\documentclass{aa}

\usepackage{natbib}
\usepackage{amsmath}
\usepackage{graphics}   
\usepackage{graphicx}   
\usepackage{epstopdf}
\usepackage{longtable}   
\usepackage{lipsum}
\usepackage[varg]{txfonts}
\usepackage{hyperref}   
\usepackage{afterpage}
\usepackage{supertabular,booktabs}
\usepackage[dvipsnames]{xcolor}
\newcommand{\Msol}{M\ensuremath{_{\sun}}}

\def\ga{\,\,\raise0.14em\hbox{$>$}\kern-0.76em\lower0.28em\hbox
{$\sim$}\,\,}
\def\la{\,\,\raise0.14em\hbox{$<$}\kern-0.76em\lower0.28em\hbox
{$\sim$}\,\,}

\begin{document}

\title{The impact of mass uncertainties on the r-process nucleosynthesis in neutron star mergers}
\author{S\'ebastien Martinet\inst{1} \and Stephane Goriely\inst{1}}

 \institute{Institut d'Astronomie et d'Astrophysique, Universit\'e Libre de Bruxelles (ULB), CP 226, B-1050 Brussels, Belgium\\
 \email: sebastien.martinet@ulb.be
 }
 \authorrunning{Martinet et al.}
 \titlerunning{Impact of nuclear uncertainties on the r-process nucleosynthesis in neutron star mergers}

\date{Accepted 20/12/2024}

\abstract
{Theoretically predicted yields of elements created by the rapid neutron capture (r-) process carry potentially large uncertainties associated with incomplete knowledge of nuclear properties \b{and} approximative hydrodynamical modeling of the matter ejection processes. One of the dominant uncertainties in determining the ejecta composition and radioactive decay heat stems from the still unknown nuclear masses of exotic neutron-rich nuclei produced during the neutron irradiation.
}
{We investigate both the model (systematic) and parameter (statistical) uncertainties affecting nuclear mass predictions and explore their impact on the r-process production, and subsequently on the composition of neutron star merger ejecta.}
{The impact of correlated model uncertainties on masses is estimated by considering five different nuclear mass models that are known to provide an accurate description of known masses. In addition, the uncorrelated uncertainties associated with local variation of model parameters are estimated using a variant of the backward-forward Monte Carlo method to constrain the parameter changes to experimentally known masses before propagating them consistently to the unknown masses of neutron-rich nuclei. 
The impact of nuclear mass uncertainties is propagated to the r-process nucleosynthesis in a 1.38-1.38~M$_{\odot}$ neutron star merger model considering a large and representative number of trajectories.} 
{We find that the uncorrelated parameter uncertainties lead to the ejected abundances uncertainties of 20\% up to $A \simeq 130$, 40\% between $A=150$ and 200, and peaks around $A \simeq 140$ and $A \simeq 203$ giving rise to deviations around 100 to 300\%.
The correlated model uncertainties remain larger than the parameter ones for most nuclei. However, both model and parameter uncertainties have an important impact on the heavy nuclei production.
}  
{Improvements to nuclear models are still crucial in reducing uncertainties in predictions related to the r-process nucleosynthesis. 
Both correlated model uncertainties and determining coherently parameter uncertainties are key in the sensitivity analysis.}

\keywords{Nuclear reactions, Nucleosynthesis,  stars:neutron}

\maketitle

\section{Introduction}

The r-process in stellar nucleosynthesis is essential for explaining the production of stable (and some long-lived radioactive) neutron-rich nuclides heavier than iron observed in stars of various metallicities, including in the Solar System \citep[see reviews of][]{Arnould07,Arnould20,Cowan21}.

Over the years, various nuclear-physics-based and astrophysics-free models of the r-process, varying in sophistication, have been developed. Advances in modeling type-II supernovae and $\gamma$-ray bursts have generated excitement around the neutrino-driven wind environment. However, achieving a successful r-process {\it ab initio} remains elusive without tuning relevant parameters (neutron excess, entropy, expansion timescale) in a manner not supported by the most sophisticated existing models \citep{Janka17}. While these scenarios hold promise, especially with their potential to contribute significantly to the Galactic enrichment, they are plagued by significant uncertainties, primarily stemming from the still incompletely understood mechanism behind supernova explosions and the persistent challenges in obtaining suitable r-process conditions in self-consistent dynamical explosion and neutron star (NS) cooling models \citep{Janka17,Wanajo18a,Bollig21}. In particular, a subclass of core-collapse supernovae, known as collapsars and corresponding to the fate of rapidly rotating and highly magnetized massive stars, often considered  as the origin of observed long $\gamma$-ray bursts, could emerge as a promising r-process site \citep{Nishimura15,Siegel19b,Just22b}. The production of r-nuclides in these events may be associated with jets predicted to accompany the explosion or with the accretion disk forming around a newly born central black hole (BH).

Since the early 2000s, special attention has been directed toward NS mergers (NSM) as potential r-process sites, especially after hydrodynamic simulations confirmed the ejection of a non-negligible amount of matter from the system. Newtonian, conformally flat general relativistic, as well as fully relativistic hydrodynamical simulations involving NS-NS and NS-BH mergers with microphysical equations of state, have shown that typically a fraction, ranging from $10^{-3}\,M_\odot$ up to more than 0.1\,M$_\odot$, can become gravitationally unbound on roughly dynamical timescales due to shock acceleration and tidal stripping. Additionally, the relic object, either a hot, transiently stable hyper-massive NS followed by a stable supermassive NS, or a BH-torus system, can lose mass through outflows driven by various mechanisms \citep{Just15}. 

Increasingly sophisticated simulations have reinforced the idea that the ejecta from NSM serve as a viable r-process sites up to the third abundance peak { (76 $\leq Z \leq$ 88)}, and the actinides { ($Z \geq $ 89)}. The enrichment of r-nuclides is predicted to originate from both the dynamical (prompt) material expelled during the NS-NS or NS-BH merger phase and the outflows generated during the post-merger remnant evolution of the relic NS-torus and BH-torus system \citep{Just15,Just23}. The resulting abundance distributions align closely with the Solar System distribution and  various elemental patterns observed in low-metallicity stars \citep{Cowan21}. Moreover, the ejected mass of r-process material, combined with the predicted astrophysical event rate (around 10\,My$^{-1}$ in the Milky Way), can account for the majority of r-material in our Galaxy { \citep[see for example][]{Shen15}}.  Another compelling piece of evidence supporting  NSM as r-nuclide producers comes from the highly important  gravitational-wave and electromagnetic observation of the kilonova GW170817 in 2017 \citep{Abbott17,Watson19,Gillanders22}. 

Despite recent successes in nucleosynthesis studies for NSM, the details of r-processing in these events is still subject to various uncertainties, particularly in the nuclear physics input \citep[see, e.g.][]{Mendoza15,Kullmann23}. One of the dominant uncertainties stems from the still unknown nuclear masses of exotic neutron-rich nuclei produced during the neutron irradiation. Indeed, in typical NSM conditions, the high neutron density and high temperatures lead the nuclear flow to be essentially constrained by radiative neutron captures and photoneutron {emissions. Despite the relatively wide literature on the impact of masses on the r-process nucleosynthesis \citep[see e.g.][]{Goriely15c,Mendoza15,Sprouse20,Mumpower16,Kullmann23}, the propagation of mass uncertainties on the composition of the ejecta remains a complicated task, especially in view of the different types of correlation embedding differential quantities, such as the neutron separation ($S_n$) or $\beta$-decay ($Q_{\beta}$) energies.}

In this paper, we study both nuclear model (or equivalently ``systematic'') and nuclear parameter (often referred to as ``statistical'') uncertainties affecting the prediction of theoretical masses. { The aim is to determine coherently their impact} on the composition of the ejecta of NSM through their influence on the neutron-capture to photoneutron rate ratios.

Section \ref{Sect:Method} presents the method to obtain model and parameter uncertainties, with a special emphasis on the application of the Backward-Forward Monte Carlo (BFMC) approach to constrain parameter uncertainties on experimental data. In Section \ref{Sect:Impact_r-process}, we study the impact of both {the parameter and model} uncertainties on the r-process nucleosynthesis in a 1.38-1.38\Msol\ NSM simulation. 
Finally in Sect. \ref{sec:conc}, we discuss the results of this work and the potential perspectives of this sensitivity study.

\section{Method}
\label{Sect:Method}
This section outlines the methodology employed to assess uncertainties in nuclear mass models and their impact on critical nuclear parameters relevant to the r-process nucleosynthesis. Understanding these uncertainties is essential for accurately predicting the behavior of nuclei far from stability and their role in astrophysical processes.

We first examine the uncertainties in various nuclear mass models and their accuracy in predicting masses of neutron-rich nuclei. Next, we use the Backward-Forward Monte Carlo (BFMC) method to quantify parameter uncertainties in neutron separation energies and $\beta$-decay energies. We then compare different methods for estimating these uncertainties and ensure consistency in predictions for neighbouring nuclei. This methodology provides a foundation for analyzing how these uncertainties impact r-process nucleosynthesis predictions.

\subsection{Nuclear model uncertainties}
\label{sec:mod}

{
Many different mass models have been proposed so far. Such models can be characterized by very different frameworks, some of them being based on the liquid-drop type of approaches with additional microscopic corrections, others on the mean-field method either using relativistic or non-relativistic energy density functionals \cite{Lunney03}. Even within a given framework, different expressions are used to determine the various ingredients and corrections to the mass calculations (e.g. a volume or surface pairing interaction, inclusion of extra terms in the functional, coordinate or oscillator basis representation, \dots). These different physical descriptions used to estimate the nuclear mass define model uncertainties.
}

{The model uncertainties are treated in a similar way as done in \citet{Kullmann23} by adopting different nuclear mass models that have proven their capacity to reproduce fairly accurately {known masses}, i.e. typically with a global root-mean-square (rms) deviation with respect to all known experimental masses below 0.8~MeV. More specifically, we adopt here { 6} different mass models, based either on the macroscopic-microscopic or the mean-field approach, namely
\begin{enumerate}
\item Skyrme-HFB model with BSkG3 interaction \citep[hereafter BSkG3,][]{Grams23}
\item Skyrme-HFB model with BSk24 interaction  \citep[hereafter HFB-24,][]{Goriely13a}, 
\item Gogny-HFB model with D1M interaction \citep[hereafter D1M,][]{Goriely09a}
\item Gogny-HFB model with the three-range D3G3M interaction \citep[hereafter D3G3M,][]{Batail24}
\item Weisz\"aker-Skyrme macroscopic-microscopic model \citep[hereafter WS4,][]{Wang14}
\item Finite-range droplet macroscopic-microscopic model \citep[hereafter FRDM12,][]{Moller16}
\end{enumerate}

In order to  meet at best the nuclear-physics needs of the r-process, we require the applied models to be both accurate with respect to experimental observables but also as reliable as possible, {\it i.e.} to be based on a physically sound model that is as close as possible to a microscopic description of the nuclear systems.
The first criterion is an objective measure, while the latter criteria may be seen as more subjective, namely that nuclear models which are as close as possible to solving the nuclear many-body problem on the scale of the nuclear chart have a stronger predictive power.
This second criterion is, however, fundamental for applications involving extrapolation away from experimentally known regions, in particular towards exotic neutron-rich nuclei of relevance for the r-process.

 The above-mentioned mass models all lead to a relatively accurate description of the experimental masses \citep{Wang21} for all the 2457 nuclei with $Z$ and $N$ larger than 8, and are all characterized by an rms deviation below typically 0.8~MeV. 
However, this overall accuracy does not imply a reliable extrapolation far away from the experimentally known region since models may achieve a small rms value by mathematically driven or unphysical corrections or can have possible shortcomings linked to the physics theory underlying the model.  In particular, despite their great empirical success, the macroscopic-microscopic approach suffers from major shortcomings, such as the incoherent link between the macroscopic part and the
microscopic correction, the instability of the mass prediction to different parameter sets, or the instability of the shell correction \citep{Pearson00,Lunney03}. For this reason, microscopic mean-field mass models have been developed for the past two decades and { are expected to provide more reliable predictions } for r-process applications. 

{ For the exotic neutron-rich nuclei relevant to the r-process, for which no experimental information is available}, model uncertainties have been shown to dominate over the parameter uncertainties \citep{Goriely14}. By definition, the model uncertainties are correlated by the underlying model, so that their propagation into r-process nucleosynthesis calculation cannot be done through standard Monte-Carlo-type of approaches. 

\subsection{Nuclear parameter uncertainties}
\label{sec:par}
\begin{figure*}
    \centering
    \includegraphics[width=0.98\textwidth]{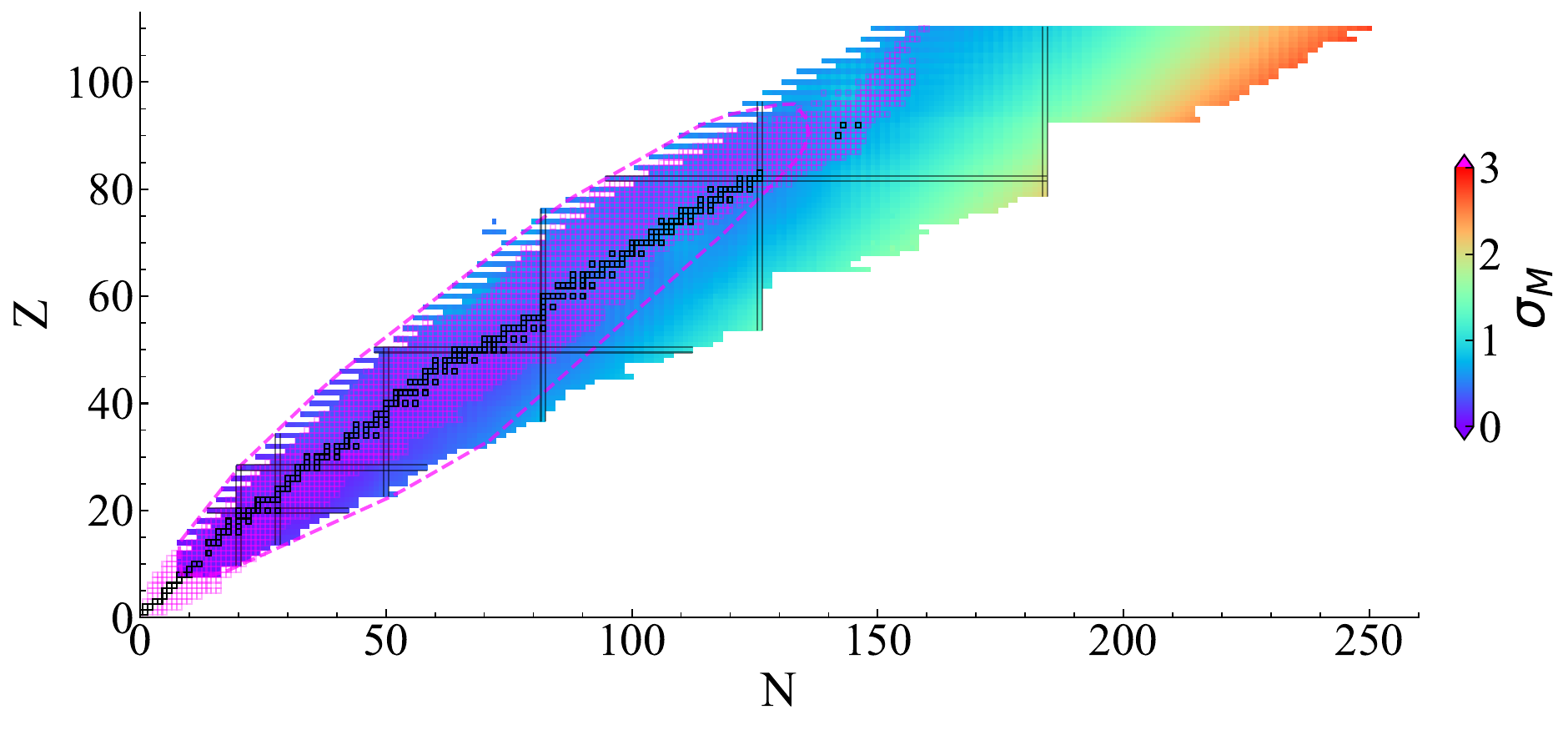}
\caption{Case 1: Representation in the ($N$,$Z$) plane of the uncorrelated masses uncertainties $\sigma_M(Z,N)$ (in MeV) obtained from 10931 runs with $\protect\sigma_{M,{\rm exp}} <0.8$~MeV. See \citet{Goriely14} for more details. The nuclei outlined with magenta open squares are the experimentally known masses. { The magenta dashed line is a contour of nuclei with $\sigma_M$ below 0.55~MeV. The shells closures are displayed in plain lines.}
} 
\label{fig:sigma_case1}
\end{figure*}
{
In contrast to model uncertainties describing the physical nature of the nuclear system, the parameter uncertainties are associated with local variations of the many parameters attached to a given model. Since none of the model is rooted in a parameter-free fundamental theory, model parameters are tuned to reproduce at best experimental data. Even if such a fit leads to an optimum description of the relevant observables, local variation of these parameters may give rather similar reproduction of the data but still lead to non-negligible differences when applied to unknown nuclei. These define the parameter uncertainties.
In comparison with model uncertainties, much less effort has been devoted to estimate the parameter uncertainties affecting masses \citep{Sprouse20}. 
}

For this study we consider the parameter uncertainties affecting { one specific mass model: }the Hartree-Fock-Bogolyubov (HFB) mass predictions, HFB-24 \citep{Goriely13a}, as studied in details in \citet{Goriely14} within the BFMC framework. { Each nuclear model has its own parameter uncertainties, however constraining them is a tedious work that needs a dedicated study \citep[see][]{Goriely14}. }{ In HFB-24, there are 30 model parameters, including 16 Skyrme, 5 pairing, 4 Wigner, and 5 collective parameters. In \citet{Goriely13a}, each parameter was adjusted to optimize the prediction of the 2353
experimental masses of nuclei with $N$ and $Z\geq 8$ available in the 2012 Atomic Mass Evaluation \citep[]{Audi2012}. In \citet{Goriely14}, 21 of these 30 parameters (corresponding to the Skyrme and pairing interactions) were varied locally around the HFB-24 minimum in order to study their impact on the extrapolated mass, as explained below.
}

{ In the study of parameter uncertainties, the range of local variation for each parameter remains to be defined and is critical to estimate the magnitude of their impact on extrapolated} masses.} For this reason, it is fundamental that such local parameter variations be constrained as much as possible by { available experimental data}, before being applied to {unknown} nuclei. In our case, it is possible to estimate the impact of such parameter uncertainties on calculated rates by propagating them by Monte Carlo (MC) sampling constrained by available experimental masses \citep{Wang21}. The method used here corresponds to the BFMC approach, as summarized below.

\subsubsection{{ The concept of the BFMC method}}
\label{sect:BFMC}

The BFMC method \citep{Chadwick07,Bauge11,Goriely14} relies on the sampling of the model parameters and the
use of a generalized $\chi^2$ estimator to quantify the likelihood of each {simulation} with respect to a given set of experimental constraints, here the experimentally known masses \citep{Wang21}. 

The backward MC is used {to select} the suitable parameter samples that agree with experimental constraints. 
{In the BFMC method, the $\chi^2$ estimator is utilized to {quantify} a likelihood function that weights a given sample of  \{$p_1$,...,$p_n$\} parameters when the $N_e$ associated
observables \{$\sigma_1$,...,$\sigma_{N_e}$\} closely match experimental data.} Experimentally constrained masses are assumed to be independent. This assumption allows us to use a $\chi^2$ criterion instead of a generalized weighting function. Thus, the weighting function is simply 1 when $\chi^2 \leq \chi_{\rm crit}^2$, and 0 otherwise, where $\chi_{\rm crit}^2$ is a chosen critical value of the $\chi^2$ { (typically the rms deviation of the model with respect to experimental masses)}. { We obtain from this backward step a subset of MC parameter combinations that are compatible with experimental constraints. }{Then, for the forward MC step, the selected subset of MC parameter combinations is applied to the calculation of the unmeasured quantities {(here the experimentally unknown masses)}.
}



\subsubsection{Parameter uncertainties associated with HFB-24 masses}

Parameter uncertainties associated with HFB-24 masses have been estimated within the BFMC method \citep{Goriely14}. {At the end of the backward MC step, a subset of $N^\xi_{\rm comb}=10931$ combinations of parameters constrained by experimental masses ($\sigma_{M,{\rm exp}} \le 0.8$~MeV) was obtained. 
For the forward MC step, this selected set of the MC parameter combinations is applied to the calculation of the $\sim 5000$ unknown masses of neutron-rich nuclei. { By definition, this method yields uncorrelated uncertainties.} It should be kept in mind that the results obtained by \citet{Goriely14} assume that the 
deformation energy is not affected by local parameter changes. 
Even if larger uncertainties could not be excluded for strongly deformed nuclei, 
such results are exact for spherical nuclei and in particular for those 
lying close to the neutron shell closures of particular interest for r-process applications. 

{ In \citet{Goriely14},
}
the uncertainties associated with such local changes of the HFB parameters were found to remain smaller than those associated with non-local changes described by the 27 different HFB mass models.  
It should, however, be kept in mind that for each mass model considered, 
uncertainties due to the parameter variations should also be applied and that if the model uncertainties are correlated by the underlying model, the parameter uncertainties remain uncorrelated. { In the present work, the parameter uncertainties are considered only for the HFB-24 mass model.}

\subsection{Approaches to determine nuclear parameter uncertainties {on $S_n$ and $Q_\beta$}}
\label{sec:cases}

In this section, we present {four possible ways to determine from the mass parameter uncertainties the uncertainties affecting $S_n$ and $Q_\beta$ values of interest to the r-process nucleosynthesis.
The separation energy $S_n$ of a ($Z$,$N$) nucleus defined as
\begin{equation}
    S_n(Z,N)=M(Z,N-1)+m_{\rm neut}-M(Z,N) \quad ,
     \label{eq:Sn}
\end{equation}
where $M$ is the atomic mass { \citep[estimated from the nuclear mass by including the masses of the electrons and their binding energies,][]{Lunney03}} and $m_{\rm neut}$ the neutron mass, plays a key role during the r-process neutron irradiation by setting the most abundant nuclei produced within an isotopic chain due to the $(n,\gamma) \rightleftharpoons (\gamma,n)$ competition (or potentially, equilibrium) \citep{Goriely92}. 
Additionally $Q_\beta$, defined as
\begin{equation}
    Q_\beta(Z,N)=M(Z,N)-M(Z+1,N-1) \quad ,
    \label{eq:Qb}
\end{equation}
is an important quantity affecting the energy production, but also the $\beta$-decay rates, { and therefore the composition of the ejecta.}
The different approaches take into account various physical considerations that impact the overall estimation of the uncertainties, in particular the correlations inherent to mass differences. 
}

Additionally, it should be stressed that to propagate nuclear uncertainties into astrophysical observables (Sec.~\ref{Sect:Impact_r-process}), a procedure similar to the one detailed in \citet{Martinet24} is followed here. This one is based on the propagation of a random choice of minimum and maximum values of $S_n$ for all nuclei of interest during the r-process irradiation, { instead of a random choice between these minimum and maximum values.} Such a procedure allows us to save computer time and still provides upper and lower values of astrophysical observables, such as the ejecta composition from NSM, that do directly depend on $S_n$ values.

{
To estimate the parameter uncertainties affecting the neutron separation energy $S_n$, we first consider two possible approaches (Case 1 and 2). Case 1 (Sect.~\ref{sec:case1}) applies the BFMC method to estimate the mass uncertainties $\sigma_M$ of all unknown nuclei and uses Eq.~\eqref{eq:Sn} to propagate those to $S_n$. In contrast, Case 2 (Sect.~\ref{Sect: Case2_theory}) applies the BFMC method directly to $S_n$, i.e. both masses defining $S_n$ values are  calculated coherently with the same parameter set selected by the BFMC method. Both Cases are detailed below.
}

\begin{figure*}
    \centering
    \includegraphics[width=0.99\textwidth]{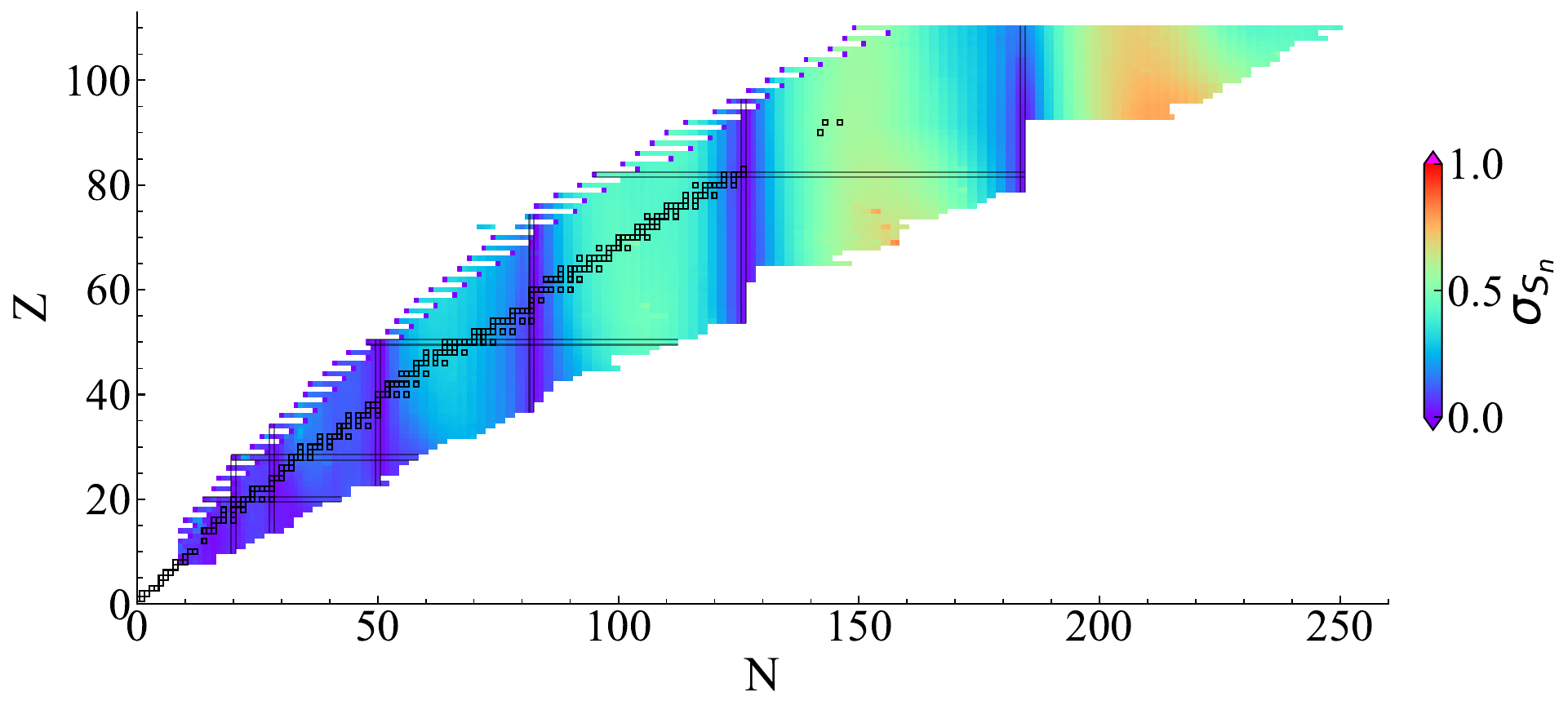}
\caption{Same as Fig.~\ref{fig:sigma_case1} for Case 2: Uncorrelated $S_n$ uncertainties (in MeV) determined in Case 2 within the BFMC method.
} 
\label{fig:sigma_case2}
\end{figure*}

\subsubsection{Case 1: $S_n$ from uncorrelated BFMC masses}
\label{sec:case1}
{ Figure \ref{fig:sigma_case1} shows the uncorrelated mass uncertainties deduced from the BFMC method \citep{Goriely14}.
} 
The nuclei outlined in magenta are the {$\sim 2500$} experimentally known masses. From the BFMC method (see Sect. \ref{sect:BFMC}), we expect the uncertainties for these nuclei to be of the order of the rms deviation of HFB-24 with respect to experimental masses, {{\it i.e.} { 0.55~MeV}}. By extension, the neighboring unknown masses have uncertainties of the same magnitude. Indeed, the combinations of parameters compatible with experimental uncertainties is not expected to induce large differences for neighboring nuclei but steadily  to increase when straying further away from the experimentally known nuclei. This effect is clearly seen in Fig. \ref{fig:sigma_case1}, where the largest uncertainties { reaching values up to $\sigma_M \simeq 3$~MeV} are found in neutron-rich nuclei close to the neutron drip line. { In turn, these uncorrelated mass uncertainties $\sigma_M$ can be used to estimate the corresponding neutron separation energies (Eq.~\ref{eq:Sn}) with their uncertainties, as illustrated in Fig.~\ref{fig:dripline}, and define our Case 1.}


\begin{figure}
    \centering
    \includegraphics[width=0.49\textwidth]{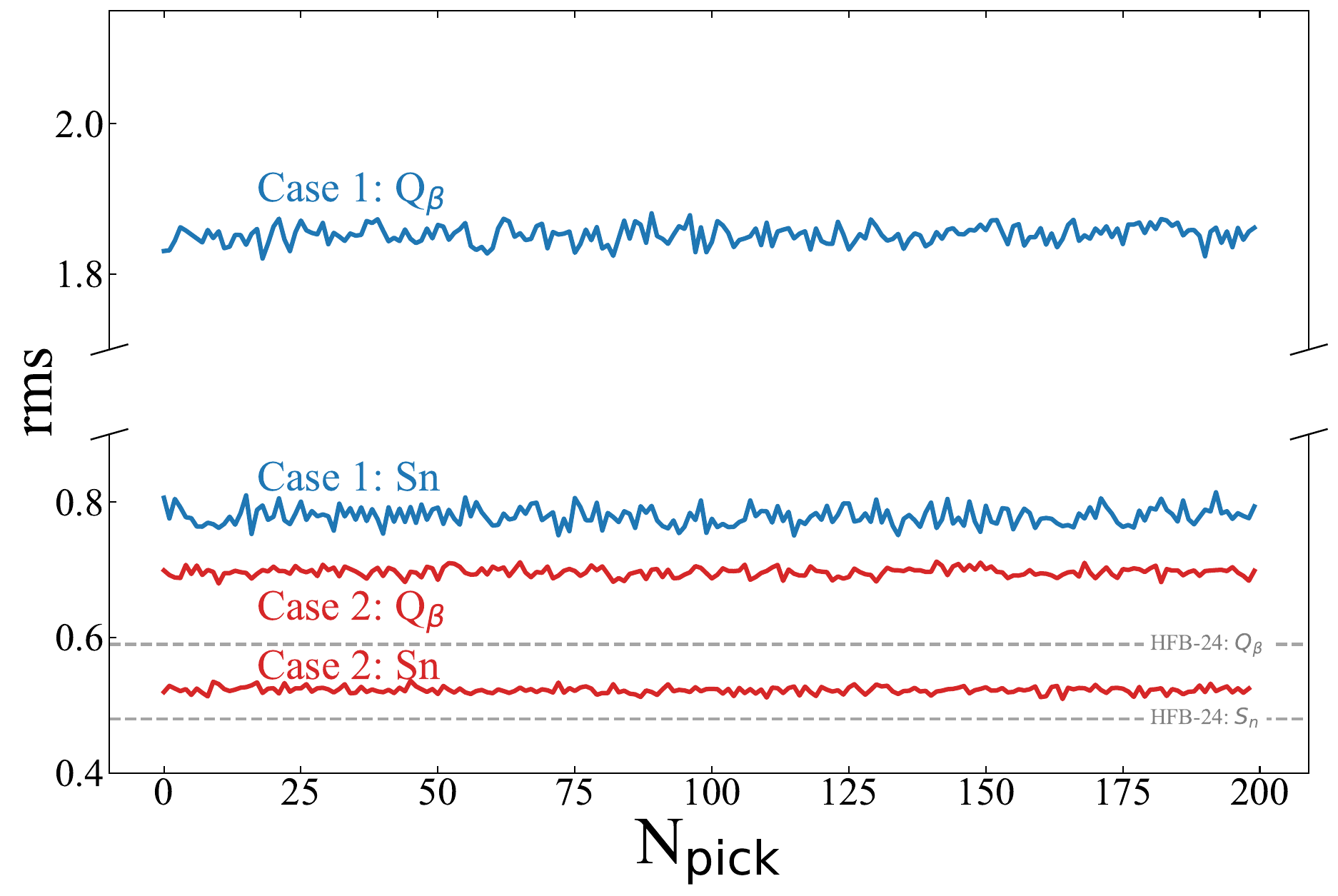}
\caption{rms values of the $S_n$ and the $Q_\beta$ when randomly picked $N_{\rm pick}$ times from Case 1 or Case 2. The rms deviation is calculated with respect to all experimentally known $S_n$ or $Q_\beta$ values \citep{Wang21}. The random pick-up is done 200 times to obtain an average rms value.
} 
\label{fig:rms}
\end{figure}
\begin{figure}
    \centering
    \hspace{-0.5cm}
    \includegraphics[width=0.48\textwidth]{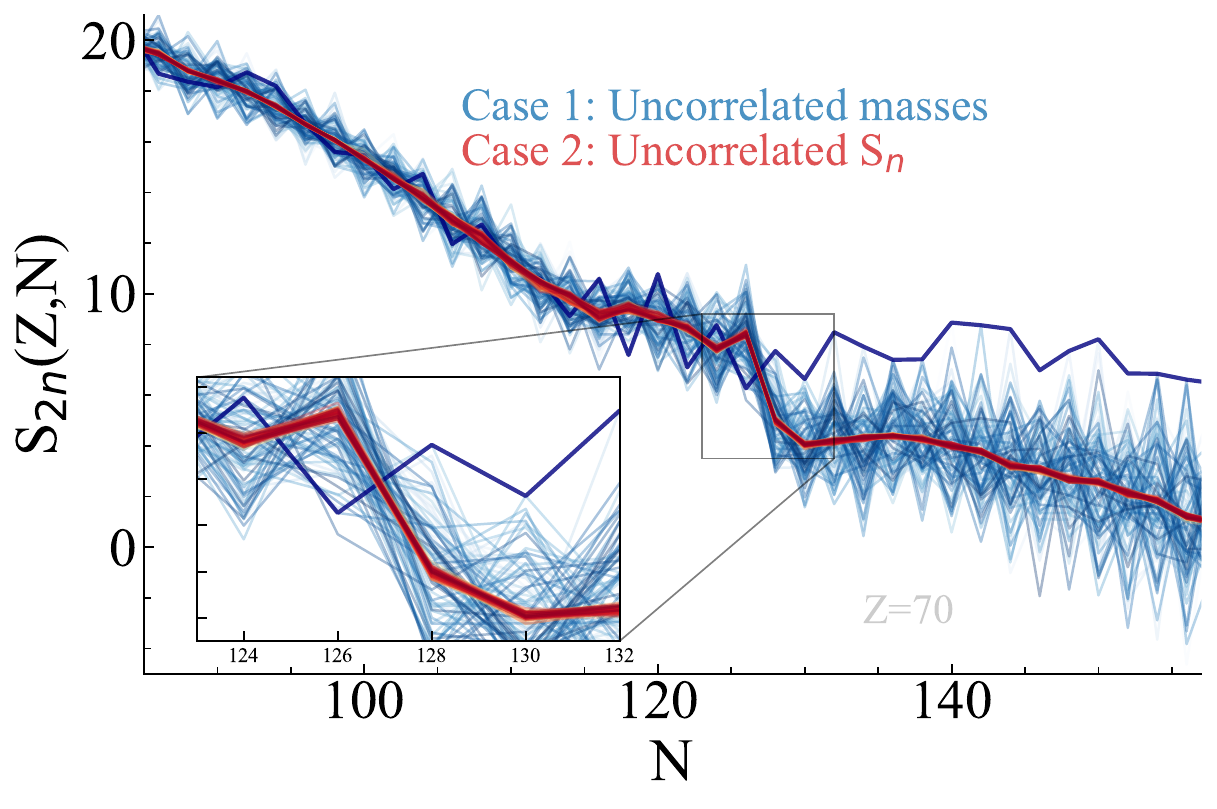}
\caption{Two neutron separation energy $S_{2n}$ for the $Z=70$ isotopic chain obtained with 200 combinations for Case 1 (blue lines), and 200 for Case 2 (red lines). A zoom on the $N=126$ shell gap shows its potential disappearance in Case 1, a clear sign of overestimating the uncertainties in this case. { One specific extreme combination of Case 1 shown by the thick blue line highlights the possible shell disappearance.} 
} 
\label{fig:S2n_no_shell_gap}
\end{figure}


\subsubsection{Case 2: Uncorrelated $S_n$ from the BFMC method}
\label{Sect: Case2_theory}
The BFMC method applied to the masses is this time used for the neutron separation energy, by computing $S_n$ for each parameter combination and then discarding those incompatible with experimental data ({\it i.e.} only combinations with $\sigma_{M,{\rm exp}} < 0.8$~MeV are retained). This approach assumes that each $S_n$, {\it i.e.} masses of both $(Z,N)$ and $(Z,N-1)$, are calculated with the same parameter set. This way we obtain a second case for our parameter uncertainties, based upon uncorrelated separation energies obtained directly from the BFMC method, { composed of a subset of $N^\xi_{\rm comb}=10643$ combinations (sets of combinations were excluded on the basis of the $S_n$ rms).}  

Figure \ref{fig:sigma_case2} shows the resulting $S_n$ uncertainties $\sigma_{S_n}(Z,N)$. In contrast to Case 1, the uncertainties are distributed non-uniformly. Indeed, two features can be seen here. First, low uncertainties in $S_n$ are found around each of the neutron magic numbers and uncertainties increase when straying further away from these areas. Second, the uncertainties globally increase when going from low $Z$/low $N$ to high $Z$/high $N$ nuclei.

The first feature is due to the separation energy being the difference between two neighbouring masses. A low uncertainty shows that { the local variation of the nuclear model parameters} does not lead to a large difference between two neighbouring nuclei. Around the magic numbers ($N= 8$, 20, 28, 50, 82, 126 and 184), a low uncertainty is found due to the conservation of the shell effects. When straying further away from the magic numbers, the uncertainty increases, especially for neutron-rich nuclei. 

{
Figure \ref{fig:dripline} compares the separation energies for each isotopic chain from $Z=21$ to $Z=100$ obtained with Cases 1 and 2. The range of $S_n$ obtained by the uncorrelated mass uncertainties from Case 1 is depicted in blue while the uncorrelated $S_n$ uncertainties from Case 2 are shown in red.
For Case 1, we can clearly see the increase of the uncertainties as we go to neutron-rich nuclei. This effect increases with increasing $N$ and $Z$. In particular, the uncertainties increase even more after the shell closures. If we look into Case 2, we see smaller uncertainties and a very mild increase in the overall uncertainties as we go to higher $Z$. It is interesting to note that Case 2 uncertainties around the neutron magic numbers are smaller than anywhere else along the isotopic chain. Moreover, we can observe that the uncertainties after the shell closures do not increase significantly in contrast to what is found in Case 1.  }

{ Figure \ref{fig:rms} shows the rms values of $S_n$ and $Q_\beta$ when choosing randomly for each nucleus a mass obtained from a random parameter set out of the 10931 BFMC combinations for Case 1 and of the 10643 BFMC combinations for Case 2. We repeat 200 times this procedure to obtain an average value of the dispersion. For Case 1 (blue curves), masses are randomly picked within the mass uncertainty $\sigma_M(Z,N)$ to compute the corresponding $S_n$ and $Q_\beta$. The HFB-24 rms deviation with respect to experimental $S_n$ amounts to 0.48~MeV and the one with respect to experimental $Q_\beta$ to 0.59~MeV, as depicted by the grey dashed lines in Fig. \ref{fig:rms}. Although all combinations obtained with the BFMC method lead to rms deviations on experimental masses smaller than 0.8~MeV, the uncorrelated $S_n$ and $Q_\beta$ of Case 1 are seen to give rise to rms values significantly higher than those obtained by HFB-24 mass model. This test clearly shows that the $S_n$ and $Q_\beta$ values deduced from uncorrelated masses are not properly constrained by experimental $S_n$ and $Q_\beta$ values, or equivalently by the underlying correlations given by Eqs.~(\ref{eq:Sn}-\ref{eq:Qb}).


To understand such differences for Case 1, Fig.~\ref{fig:S2n_no_shell_gap} shows the two neutron separation energy $S_{2n}$ for one isotopic chain (here $Z=70$), using the same 200 random combinations discussed in Fig. \ref{fig:rms}. A much larger dispersion of $S_{2n}$ can be observed for Case 1 in comparison with Case 2. The important feature is highlighted in the insert, where we can see that while a clear shell gap at $N=126$ is maintained for all combinations of Case 2, a potential disappearance of this shell effect is found for some combinations of Case 1, as highlighted by the thick blue line. The shell gap at neutron magic numbers is a well-known physical effect, so that parameter combinations that would lead to the artificial disappearance of these shell effects need to be questioned.  
}

Such a feature found for most isotopic chains, combined with Fig. \ref{fig:rms}, underlines the overestimation of parameter uncertainties when estimating $S_n$ from uncorrelated masses (Case 1). For this reason, we advise not to use Case 1 to estimate parameter uncertainties affecting $S_n$. Case 2 considers such correlations, but may underestimate the uncertainties since the calculation of both associated masses are calculated with the same parameter set. Anti-correlation between neighboring $S_n$ values may also not be guaranteed in Case 2. In addition, a given set of $S_n$ values may not correspond to one unique consistent set of masses. { Both these issues to further improved Case 2 are tackled below with Cases 3 and 4, respectively.}


\subsubsection{Case 3: anti-correlation between $S_n$}


The neutron separation energy, by definition (Eq.~\ref{eq:Sn}), links the atomic mass $M(Z,N)$ to the mass of the neighbor nucleus $M(Z,N-1)$. 
Figure \ref{fig:Sn_vs_Sn_Nm1} shows the density heatmap of the parameter uncertainties of the separation energy of a ($Z$,$N$) nucleus obtained in Case 2 as a function of the uncertainties affecting the ($Z$,$N-1$) one for $Z=30$ to $Z=101$ isotopic chains. As can be seen for all isotopic chains, an anti-correlation is clear between two neighbouring $S_n$. This comes from the fact that neighbouring $S_n$ are linked by the uncertainty of the mass of one of the two nuclei. A maximum uncertainty on $S_n(Z,N)$ imposes a minimum ($Z$,$N$) mass or a maximum ($Z$,$N$-1) mass. This in turn will result in the $S_n(Z,N-1)$ being minimal. 

To take this anti-correlation into account, we construct Case 3 from Case 2, for which we impose that each { even-$N$} $S_n(Z,N)$ be anti-correlated to its ($Z$,$N-1$) neighbouring value. In such conditions, we ensure that parameter uncertainties for ($Z$,$N-1$) be coherent with its ($Z$,$N$) neighbour.  

One of the main caveats of Case 3 is that by keeping a random pick of the uncertainties for the { even-$N$} nuclei and imposing anti-correlation with their { odd-$N$} ($Z$,$N-1$) neighbour nuclei only, it is possible that different masses are used for the same nucleus. 
\begin{figure}
    \centering
    \includegraphics[width=0.48\textwidth]{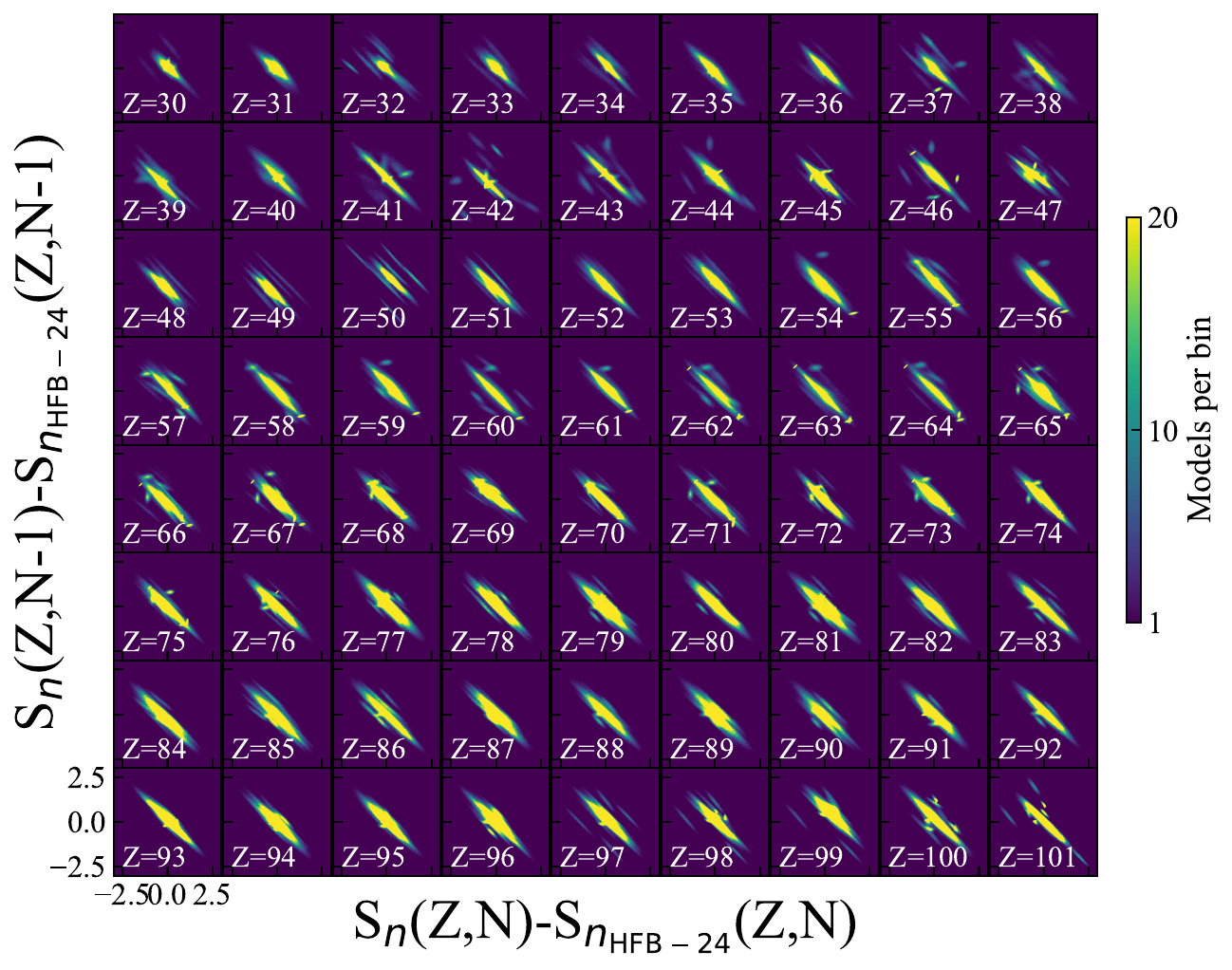}
\caption{{Heatmap representing the parameter uncertainty of $S_n(Z,N-1)$ (relative to HFB-24 values) as a function of $S_n(Z,N)$ uncertainties (relative to HFB-24 values) { for each BFMC combinations selected in Case 2 (10643 combinations). The subplots are using the same axis and are separated by isotopic chain to show the correlation for each of them}. Each frame is 500 per 500 bins, and each bin is color-coded to the number of { combinations} in it.  
}  
} 
\label{fig:Sn_vs_Sn_Nm1}
\end{figure}

\subsubsection{Case 4: Determining uncertainties of $S_n$ and $Q_\beta$ with consistent masses}
\label{sect:Sn_correlation}

{To cure for the incoherent approach underlying Case 3, we now consider a Case 4 where we ensure that the neutron separation energies obtained with Case 2, not only lead to anti-correlation between neighbouring nuclei, but also that each set of $S_n$ correspond to a coherent set of masses. In addition, this Case 4 includes not only $S_n$ but also $Q_\beta$ as the main quantities to be minimized or maximized in the uncertainty analysis and their propagation. 
}

{In the same way as discussed for neighboring $S_n$ values, $Q_\beta(Z,N)$ given by Eq.~(\ref{eq:Qb}) should be anti-correlated with $Q_\beta(Z+1,N-1)$. Moreover, since $S_n(Z,N)$ and $Q_\beta(Z,N)$ are linked by the same ($Z$,$N$) mass, they should also be anti-correlated.
Similarly to the results shown in Fig. \ref{fig:Sn_vs_Sn_Nm1}, the density heatmap of the parameter uncertainties of the $Q_\beta(Z+1,N-1)$ versus those of $Q_\beta(Z,N)$ presents the same behavior. The anti-correlation also propagates along the isobaric chain, correlating the uncertainties of each { nucleus composing it}. 
Similar density heatmaps are also found for the anti-correlation between $Q_\beta(Z,N)$ and $S_n(Z,N)$ uncertainties. 
}


In Fig.~\ref{fig:Schema_Sn}, we show how imposing these 3 anti-correlations can result in { imposing} the choice of the masses throughout the whole nuclear network 
For example, on the one hand, maximizing $S_n(Z,N)$ is equivalent to choosing the minimum $M(Z,N)$ and the maximum $M(Z,N-1)$ (see Eq. \ref{eq:Sn}). On the other hand, to maximize $Q_\beta(Z,N)$ is equivalent to choosing the maximum $M(Z,N)$ and the minimum $M(Z+1,N-1)$. This lead to an anti-correlation between $S_n$ and $Q_\beta$, where $Q_\beta$ will be minimum when $S_n$ is maximum, and vice-versa.
As depicted in Fig. \ref{fig:Schema_Sn}, there will then be two resulting combinations of masses that will either maximize or minimize the $S_n$ and the $Q_\beta$, with either all maximum masses on even $N$ nuclei and minimum on odd $N$, or the opposite. In the end, we obtain 2 times 2 combinations, as we can either start { maximizing/minimizing} from the first even nuclei, and propagate the coherent masses on the isotopic chain, or we can do the same starting by the first odd nucleus.
Note that another potential case could be obtained following a similar methodology but through $Q_\beta$ along the isobaric chains.


To compute these coherent $S_n$ for Case 4, we need to find the mass of the $(Z,N-1)$ such as it maximizes $S_n(Z,N)$ keeping a coherent set of masses for the entire network. 
We can derive the coherent mass of the $(Z,N-1)$ by extracting an associated uncertainty $\sigma^*_M(Z,N-1)$ from Eq. (\ref{eq:Sn}). In the case $S_n(Z,N)$ is maximized, $\sigma^*_M(Z,N-1)$ reads
\begin{equation}
\sigma_M^*(Z,N-1)=\sigma_{S_n}(Z,N)-\sigma_M(Z,N),
    \label{eq:sigma_m*_sn}
\end{equation}
where $\sigma_M(Z,N)$ and $\sigma_{S_n}(Z,N)$ are the uncorrelated mass and $S_n$ uncertainties from the BFMC method, respectively (as extracted in Case 1 and Case 2).
Similarly, using Eq. (\ref{eq:Qb}), a new uncertainty for $M(Z+1,N-1)$ can be obtained by minimizing the $Q_\beta$ and reads
\begin{equation}
\sigma_M^*(Z+1,N-1)=\sigma_{Q_\beta}(Z,N)-\sigma_M(Z,N),
    \label{eq:sigma_m*_qb}
\end{equation}
where $\sigma_{Q_\beta}(Z,N)$ is the uncorrelated $Q_\beta$ uncertainty from the BFMC method (determined in a way similar to Case 2 for $S_n$).
By deriving this way all the upper and lower limits of the masses for the network, we can compute the corresponding $S_n$ with the assurance that both anti-correlations in $S_n$ and $Q_\beta$ and a coherent set of masses are taken into account. This approach defines our Case 4 and the corresponding 4 combinations of anticorrelated minima and maxima in the neutron separation energies.

\begin{figure}
    \centering
    \includegraphics[width=0.48\textwidth]{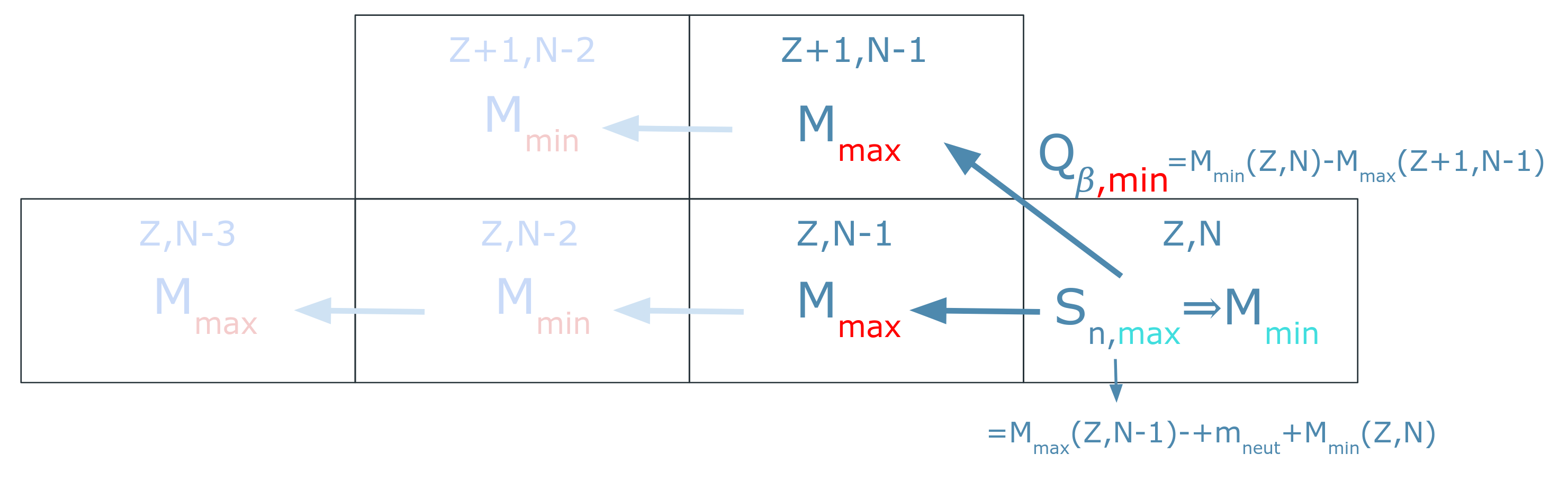}
\caption{Schematic view of the anti-correlations imposed by the $S_n$ and the $Q_\beta$ on the choices of mass uncertainties. When maximizing $S_n$($Z$,$N$), we impose $M_{\rm min}$($Z$,$N$). The consequences of this choice are depicted in red, where the mass of the ($Z$,$N-1$) will need to be minimum and the mass of the ($Z$+1,$N-1$) maximum due to the anti-correlation with $Q_\beta$.
} 
\label{fig:Schema_Sn}
\end{figure}

\section{Impact on the r-process nucleosynthesis}
\label{Sect:Impact_r-process}

\subsection{NSM simulations}




We consider here the hydrodynamical simulation of a symmetric 1.38-1.38\Msol\ binary NS system and its post-processing nucleosynthesis \citep{Just23}. The simulation includes consistently the dynamical component, the NS-torus transient and the BH-torus remnant, in one unique long-term end-to-end evolution model (up to 100~s) with intermediate remnant lifetimes (between $\sim \! 0.1\text{--}1\,$s). The model considered here corresponds to the symmetrical simulation sym-n1a6 of \citet{Just23}.

Note that the nucleosynthesis abundances are calculated using a smaller { representative} subset of trajectories to reduce the computational demand. The subset of trajectories has been chosen in such a way that they contain 10\% percent of the total mass { \citep[as discussed in][]{Kullmann23}} while still satisfactorily reproducing the properties ({\it i.e.} mass fraction distributions and heating rates) of the complete trajectory set. As a result, all calculations concerning the sym-n1a6 model include 358 trajectories. 

The nucleosynthesis is followed by a reaction network consisting of $\sim 5000$ species (depending on the set of nuclear masses adopted), { ranging from proton numbers $Z=1$ to $Z=100$} and including all isotopes from the valley of $\beta$-stability to the neutron drip line. {Elements with $Z>100$ are assumed to fission spontaneously with very short lifetimes, since their production is found to be insignificant for the conditions found in the sym-n1a6 hydrodynamical model \citep{Just23} and in view of the HFB-14 fission barriers \citep{Goriely07a,Goriely15c,Lemaitre21,Kullmann23} adopted in our present calculations}. Note that the impact of this upper $Z$ limit has been tested and found to be negligible while significantly reducing the CPU time. {All reactions and decay modes of relevance, {\it i.e.} neutron, proton and $\alpha$ capture reactions, photodisintegrations, $\beta$-decays, $\beta$-delayed processes and $\alpha$-decay are included.} If the network reaches trans-Pb species, we also include neutron-induced, spontaneous and $\beta$-delayed fission and their corresponding fission fragment distributions for all nuclei. Each fissioning parent is linked to the daughter nucleus, and the emitted neutrons may be recaptured by the nuclei present in the environment. 

Whenever experimental reaction or decay data is available, it is included in the network and are taken from the NETGEN{ \footnote{\url{http://astro.ulb.ac.be/Netgen/}}} library \citep{Xu13}, which includes the latest compilations of experimentally determined rates. 
{Whenever experimental reaction model ingredients (such as masses, as well as energies, spin and parities of ground and excited states)} are available, they are also considered in the theoretical modelling. In this case,  they also constrain the possible range of variation of the model parameters and reduce the impact on the model as well as parameter uncertainties. The model uncertainties affecting theoretical nuclear masses are propagated to the calculation of the radiative neutron capture and photoneutron rates. However, in the case of parameter uncertainties, the same radiative neutron capture rates, {as obtained on the basis of the HFB-24 mass model, are calculated} and only the photoneutron rates are estimated on the basis of the modified neutron separation energies $S_n$. Newly calculated rates are then consistently applied to the nucleosynthesis simulations to estimate their impact on the r-process yields. {Note that the uncertainties affecting masses are propagated into the $Q_\beta$-values, which therefore may affect the decay heat, but not into the $\beta$-decay rates. Their impact on the abundance distribution is found to be relatively small with respect to those stemming from changes in the neutron separation energies. }

To explore the impact of the correlated model and uncorrelated parameter uncertainties, a large number of nucleosynthesis calculations is performed, as described below.

\subsection{Propagating parameter uncertainties}
\begin{figure}
    \centering
    \includegraphics[width=0.46\textwidth]{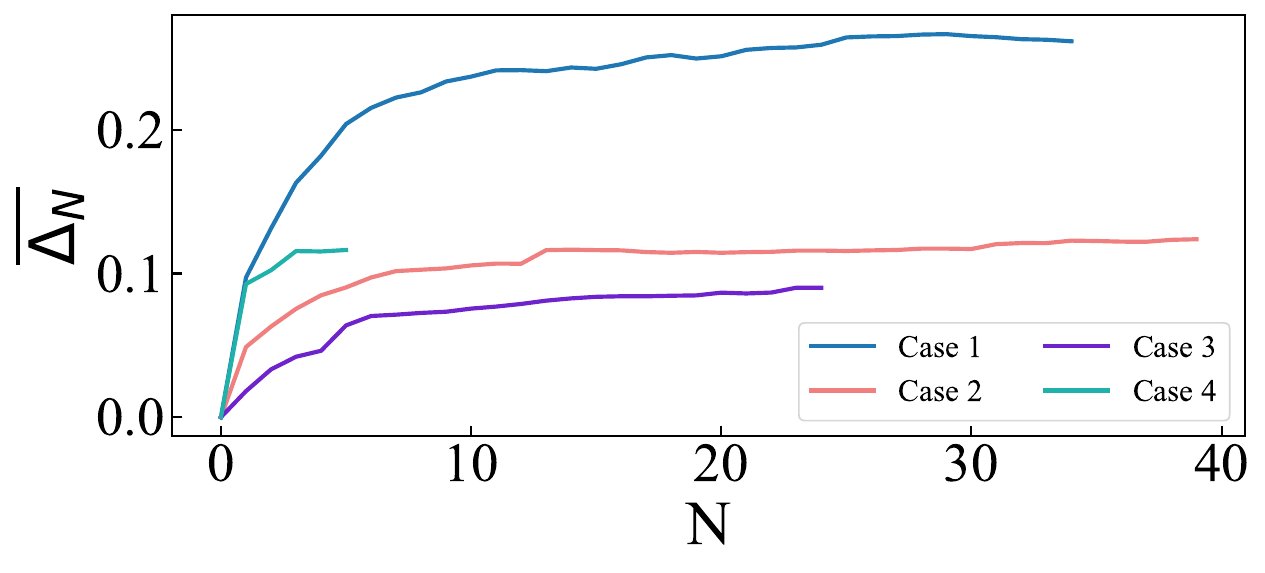}
\caption{Convergence of the averaged uncertainties $\overline{\Delta_N}$ as a function of the number $N$ of { r-process} simulations for the 4 cases considered here.} 
\label{fig:conv}
\end{figure}

\begin{figure*}
    \centering
    \includegraphics[width=0.98\textwidth]{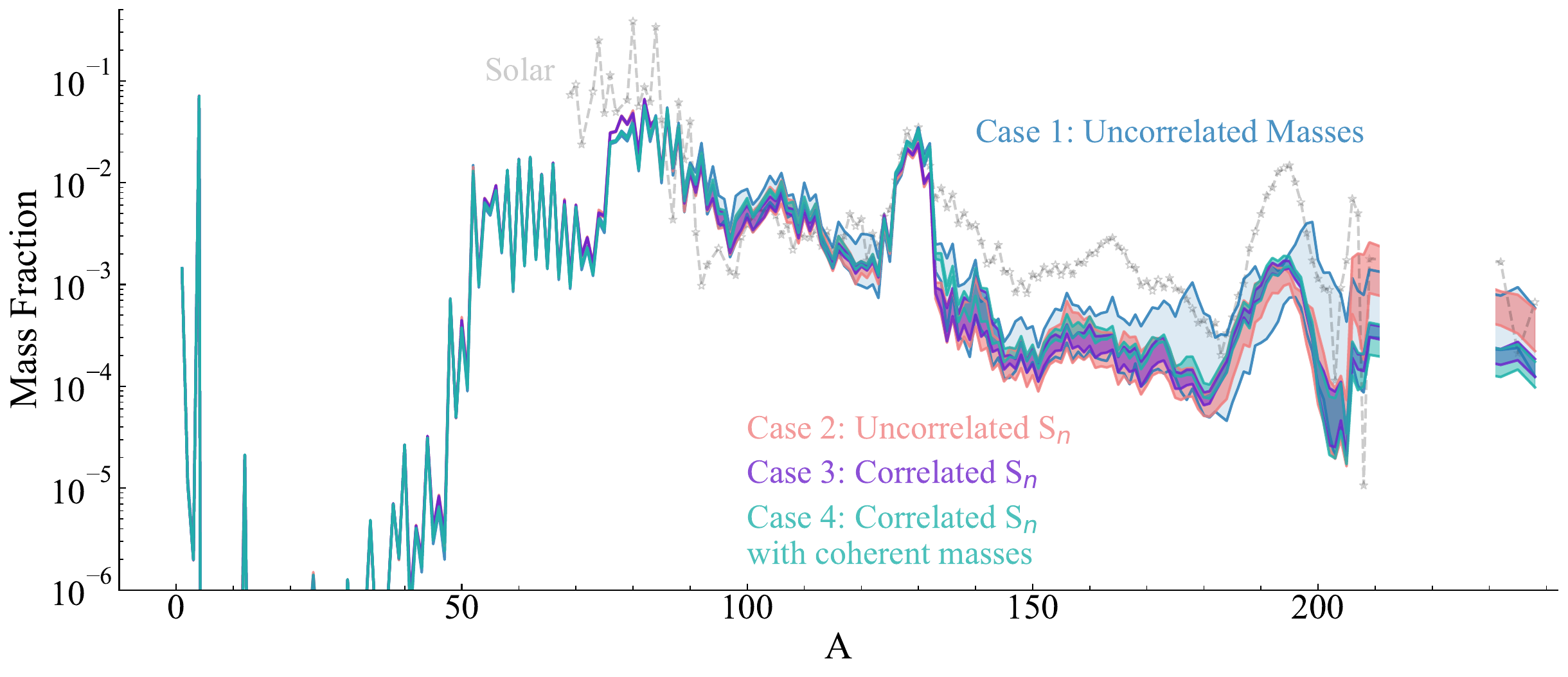}
\caption{Impact of the nuclear mass uncertainties obtained within Case 1 to 4 on the mass fractions of stable nuclei (and long-lived Th and U) of the material ejected from the binary 1.38-1.38~M$_\odot$ NSM model as a function of the atomic mass $A$. Solar r-abundances \citep{Goriely99} are shown in grey as a reference and are scaled to the peak value of the $A=130$ abundance in the Case 1 simulation.} 
\label{fig:Case_all}
\end{figure*}
To propagate the uncorrelated parameter uncertainties, a large number $N$ of sets containing randomly chosen minimum and maximum rates for each of the thousands of HFB-24 masses is produced. These are based on the maximum and minimum masses or $S_n$ obtained in Sect.~\ref{sect:BFMC} from the BFMC method and one of the 4 cases described in Sect.~\ref{sec:cases}. These random nuclear sets are then used in our NSM model to assess the impact of the parameter uncertainties on the final composition of ejecta.

One of the difficulties with randomly produced sets of masses or $S_n$ is to ensure that enough draws have been made to be representative of the full range of possible outcomes. We computed an increasing number of simulations until convergence of the upper and lower limits of the ejecta abundances is reached.
More specifically, the convergence is evaluated by the quantity
\begin{equation}
\overline{\Delta_N} = \frac{1}{N_A}\sum_{j=1}^{N_A} \Delta_{N}
\label{eq:del}
\end{equation}
where $N_A$ is the total number of nuclides and
\begin{equation}
\Delta_N = {\rm max}[\log(X)] - {\rm min}[\log(X)]
\label{eq:ddel}
\end{equation}
where $X$ is the mass fraction and $N$ is the number of r-process simulations considered (up to 50).

Figure~\ref{fig:conv} shows the convergence of the averaged uncertainties $\overline{\Delta_N}$ as a function of the number $N$ of simulations for our 4 cases. For small values of $N$, a rapid increase is expected due to the random nature of the draws, leading to large variations in the uncertainties. For more than typically $N=20$ simulations, a plateau is found and the global abundance uncertainty does not evolve anymore when compared to the value for $N=40$ simulations. 
Hence, we consider that we have convincingly converged to the total propagation of parameter uncertainties by computing $N \ge 20$ stellar simulations with as many different nuclear masses, $S_n$ or anti-correlated $S_n$ sets. Case 4 is, per construction, a set made of 4 combinations and is just indicated for completeness.     

\subsection{Case 1 vs Case 2: Uncorrelated Masses vs uncorrelated $S_n$}

Figure \ref{fig:Case_all} shows the propagated nuclear uncertainties to the final abundances of our NSM model. The solar r-abundances are depicted in grey as a reference and scaled to the peak value of the $A=130$ abundance {given by the Case 1 simulation}. Since an (n,$\gamma$)--($\gamma$,n) equilibrium is established in most of the ejecta trajectories, masses play a key role in defining where such an equilibrium is achieved in each isotopic chain. As expected from Sect~\ref{Sect: Case2_theory}, the propagation of larger $S_n$ uncertainties, especially for neutron-rich nuclei, leads to larger uncertainties in the r-process production. For the elements up to Ni, their production is mostly happening in trajectories with {a relatively high electron fraction $Y_e\ga 0.4$ \citep{Just23}. For this reason, the nucleosynthesis of $A \la 70$ nuclei is not significantly affected by mass uncertainties. In contrast, heavier species are produced by low-$Y_e$ trajectories which include the production of exotic neutron-rich nuclei for which mass uncertainties are large, in particular if obtained within Case 1. Large deviations are found in particular for $A>140$ mass fractions in this Case 1. Such deviations are already significantly reduced when adopting Case 2 uncertainties. It is important to note that Case 2 is not just a subset of Case 1. For this reason, some abundances found within Case 2 cannot be obtained within Case 1, e.g. around $A \simeq 210$.   
The possible disappearance of the $N=126$ shell closure in the overestimated uncertainties of Case 1 are translated into a third r-process peak that is either flattened or reduced.
}

\subsection{Case 3: Correlated $S_n$}
Sect. \ref{sect:Sn_correlation} discussed the anti-correlation existing between the separation energy $S_n$ of the ($Z$,$N$) nucleus and its ($Z$,$N-1$) neighbour. To take this anti-correlation into account, we now draw only randomly the maximum/minium uncertainty for the even nuclei and impose that the $S_n$ uncertainty of its ($Z$,$N-1$) neighbour take the opposite sign (\textit{i.e.} minimum/maximum uncertainty respectively). 
Figure \ref{fig:Case_all} compares Case 3 against Cases 1 and 2. As expected, { the anti-correlation forces fewer combinations which also leads to reduced uncertainties for the r-process abundances}. This effect becomes important for $A > 150$, especially for the actinide production. The discrepancy between Cases 2 and 3 shows that above $A = 205$, some extreme abundances can only be obtained for specific combinations.
It underlines that the masses are highly sensitive to nuclear uncertainties in this range and very specific combinations of the uncertainties can lead to major differences in the nucleosynthesis predictions. 

\begin{figure}
    \centering
    \includegraphics[width=0.48\textwidth]{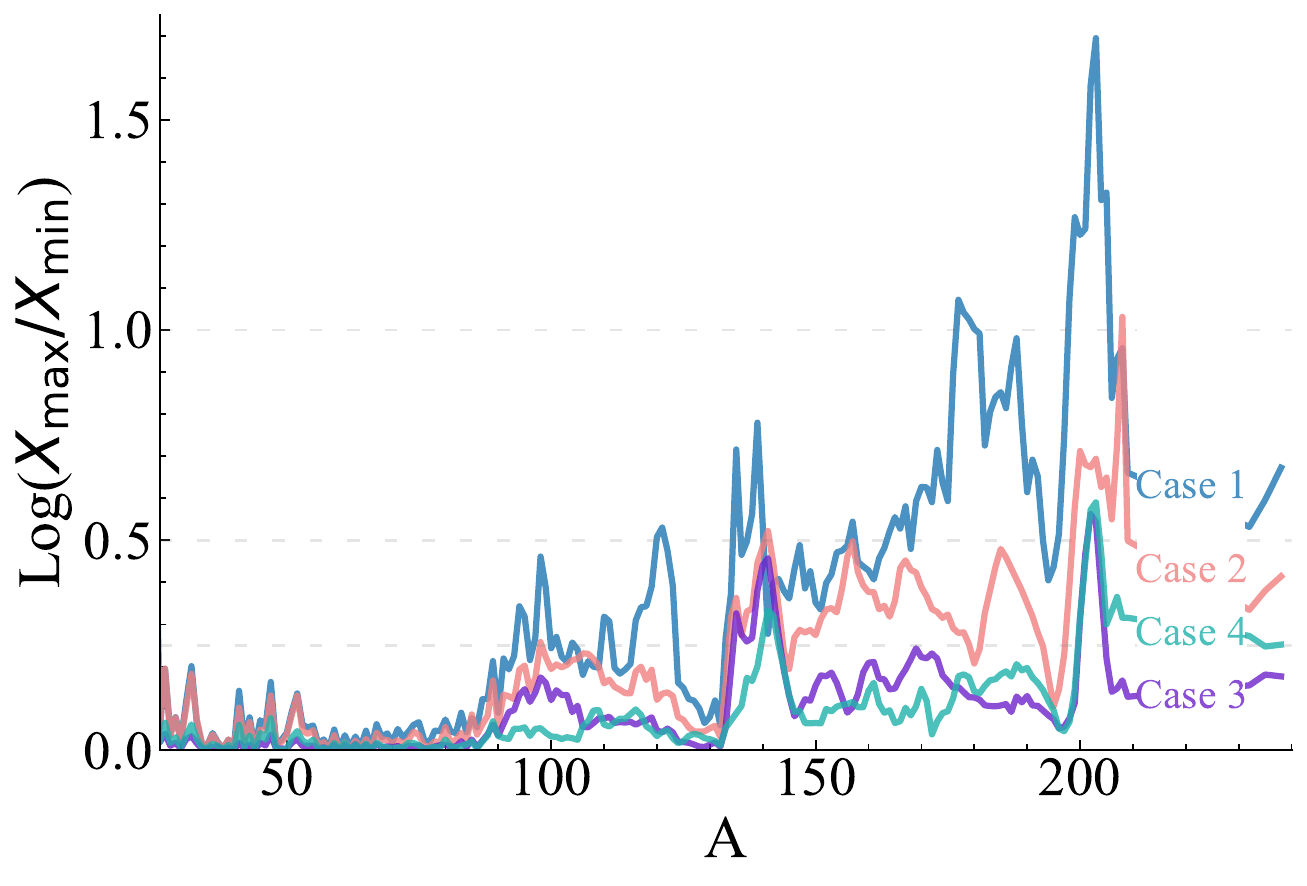}
\caption{
NSM abundances uncertainties (in log scale) given by the ratio of maximum to minimum mass fractions ($X_{\rm max}/X_{\rm min}$) resulting from $S_n$ uncertainties as obtained within Cases 1 to 4.
} 
\label{fig:abond_uncertainties_square}
\end{figure}

\subsection{Case 4: Correlated $S_n$ and coherent masses}
To tackle the incoherent approach underlying Case 3, we now consider Case 4 where we ensure that each set of $S_n$ correspond to a consistent set of masses (see Sect. \ref{sect:Sn_correlation}), and includes not only $S_n$ but also $Q_\beta$ as the main quantities to be minimized or maximized in the uncertainty analysis and their propagation. 
The resulting abundance uncertainties of Case 4 shown in Fig.\ref{fig:Case_all} are slightly smaller than for Case 3 and lead to slightly different values. However, the feature of a small production of Pb and actinides remains.  

\begin{figure*}
    \centering
    \includegraphics[width=0.98\textwidth]{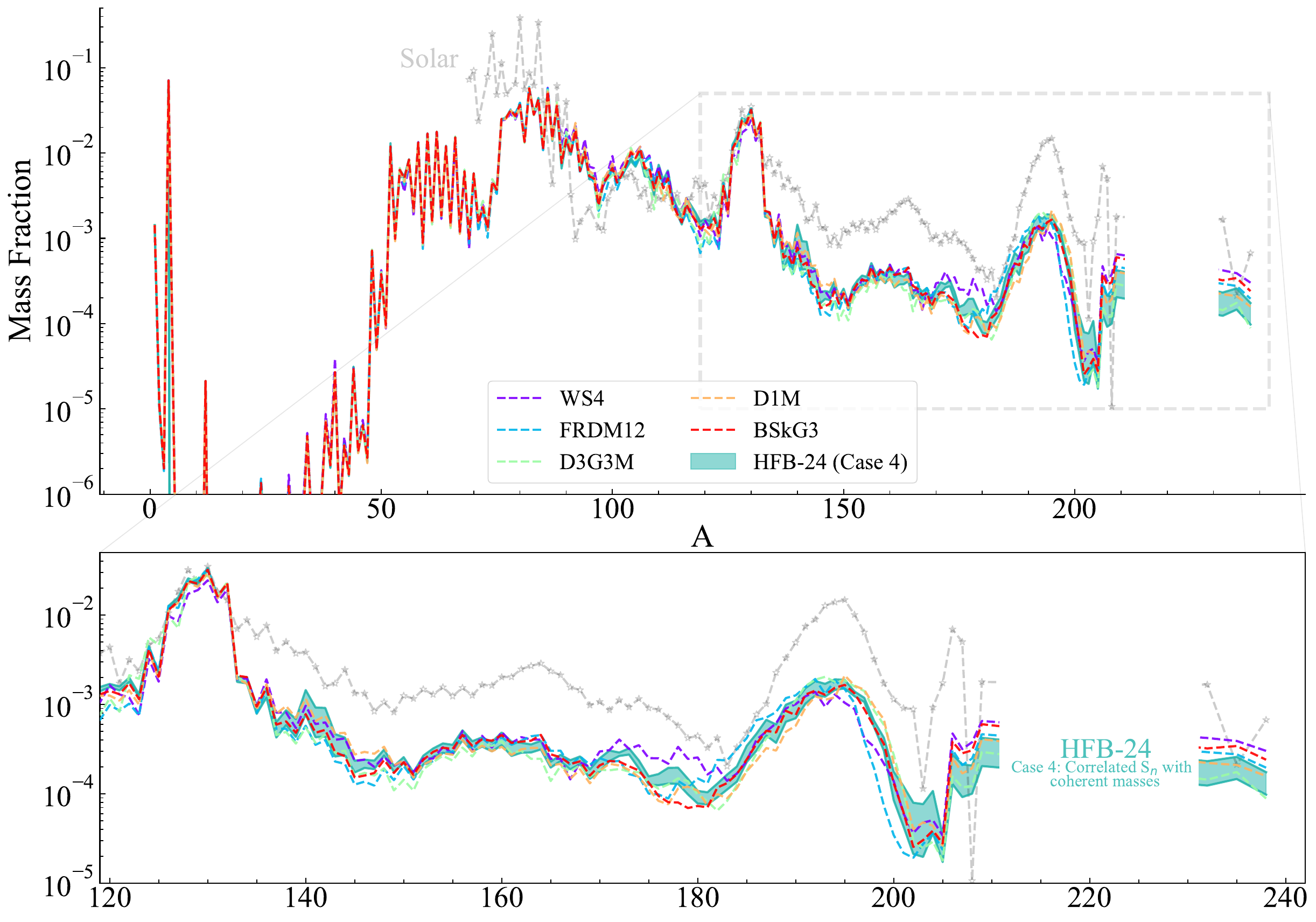}
\caption{Same as Fig.~\ref{fig:Case_all} for the impact of model vs parameter mass uncertainties. The 6 nuclear models used here are described in Sect. \ref{sec:mod}. The insert zoomed in the lower panel shows the $A > 120$ range where the r-process nucleosynthesis is dominantly impacted by the parameter and model uncertainties. The blue shaded range gives the HFB-24 associated parameter uncertainties. } 
\label{fig:Model_nuc}
\end{figure*}

\subsection{Summary of the 4 Cases}

Figure \ref{fig:abond_uncertainties_square} shows the abundance uncertainties (given by the ratio of maximum to minimum mass fractions) stemming from nuclear mass uncertainties, as described by our 4 cases. As discussed in Sec.~\ref{sec:case1}, Case 1 systematically overestimates uncertainties, in particular in the vicinity of shell closures, and should be avoided in estimating nuclear mass uncertainties. Alternatively, Case 2, while being a more reasonable estimate, remains non-physical since the anti-correlation between neighbouring $S_n$ or $Q_\beta$'s is neglected. This anti-correlation is cured by Case 3, but with the drawback that $S_n$ values may not correspond to one unique and coherent set of masses. For these reasons, both Case 2 and Case 3 should not be used to estimate and propagate nuclear uncertainties. Finally, Case 4 represents our recommended estimate of parameter uncertainties, as it fulfills all the above physical requirements. After propagation, we find with Case 4 abundance uncertainties ranging between a factor of 1.5 and 3 while Case 1 leads to deviation ranging from a factor of 3 to more than 10 (see Fig.~\ref{fig:abond_uncertainties_square}).


\subsection{Systematic vs Statistical uncertainties}

\begin{figure}
    \centering
    \includegraphics[width=0.48\textwidth]{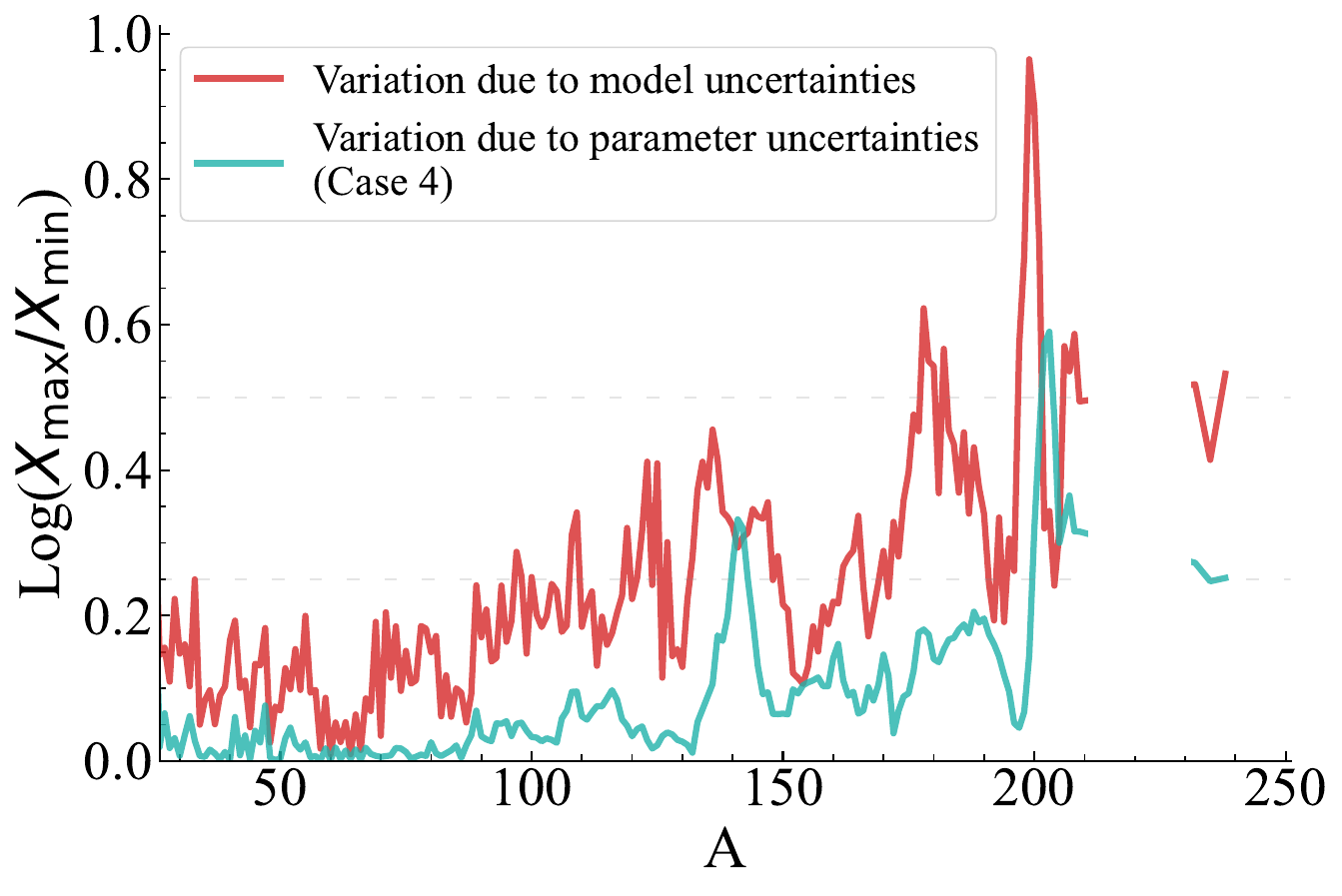}
\caption{{Abundances uncertainties ($X_{\rm max}/X_{\rm min}$) for Case 4 vs model uncertainties (using WS4, D1M, FRDM12, BSkG3, D3G3M and HFB-24 models). Here the model uncertainties are represented as if they would be uncorrelated, a common misrepresentation of the correlated model uncertainties \citep{Martinet24}.
}
} 
\label{fig:abond_uncertainties_square_misused}
\end{figure}

We now compare in Fig.~\ref{fig:Model_nuc} the impact of the parameter uncertainties affecting the HFB-24 mass model (within the Case 4 approach) with the impact of model uncertainties for which the 6 different mass models described in Sec.~\ref{sec:mod} are adopted, namely two macroscopic-microscopic models, WS4 \citep{Wang14} and FRDM12 \citep{Moller16} and 4 mean-field models, HFB-24 \citep{Goriely13a}, BSkG3 \citep{Grams23}, D1M \citep{Goriely09b} and D3G3M \citep{Batail24}. It should be recalled that each mass model has its own parameter uncertainties that should be assessed using the same methodology, as detailed in the present study. We can see that the final composition of the NSM ejecta is less affected by the HFB-24 parameter uncertainty in comparison to the impact the model uncertainties can have. For $A \ga 203$, the parameter uncertainties may however become more significant than the model ones.

{
Figure \ref{fig:abond_uncertainties_square_misused} compares the maximum-to-minimum abundance ratios obtained with Case 4 uncorrelated parameter uncertainties and those obtained from model uncertainties. It should be stressed that model uncertainties are correlated by the model \citep[see discussion in ][]{Martinet24}. In this case, the upper and lower limits should not be seen as a possible band defining minimum or maximum abundances. Comparing them to Case 4 parameter uncertainties (that are uncorrelated), we can see a clear overestimate of the uncertainties when misinterpreting the correlation underlying model uncertainties. 
The peak of Case 4 uncertainties around $A \simeq 140$ combined with the high uncertainty among models underlines a large sensitivity of the predicted abundances with respect to remaining parameter uncertainties. Interestingly, the parameter uncertainty at $A \simeq 202$ is higher than the model uncertainty, stressing also a high sensitivity with respect to parameter uncertainties.
}

{ 

\subsection{Propagation of the mass uncertainties on the neutron capture rates and r-process abundances}

\begin{figure*}
    \centering    
    \includegraphics[width=0.98\textwidth]{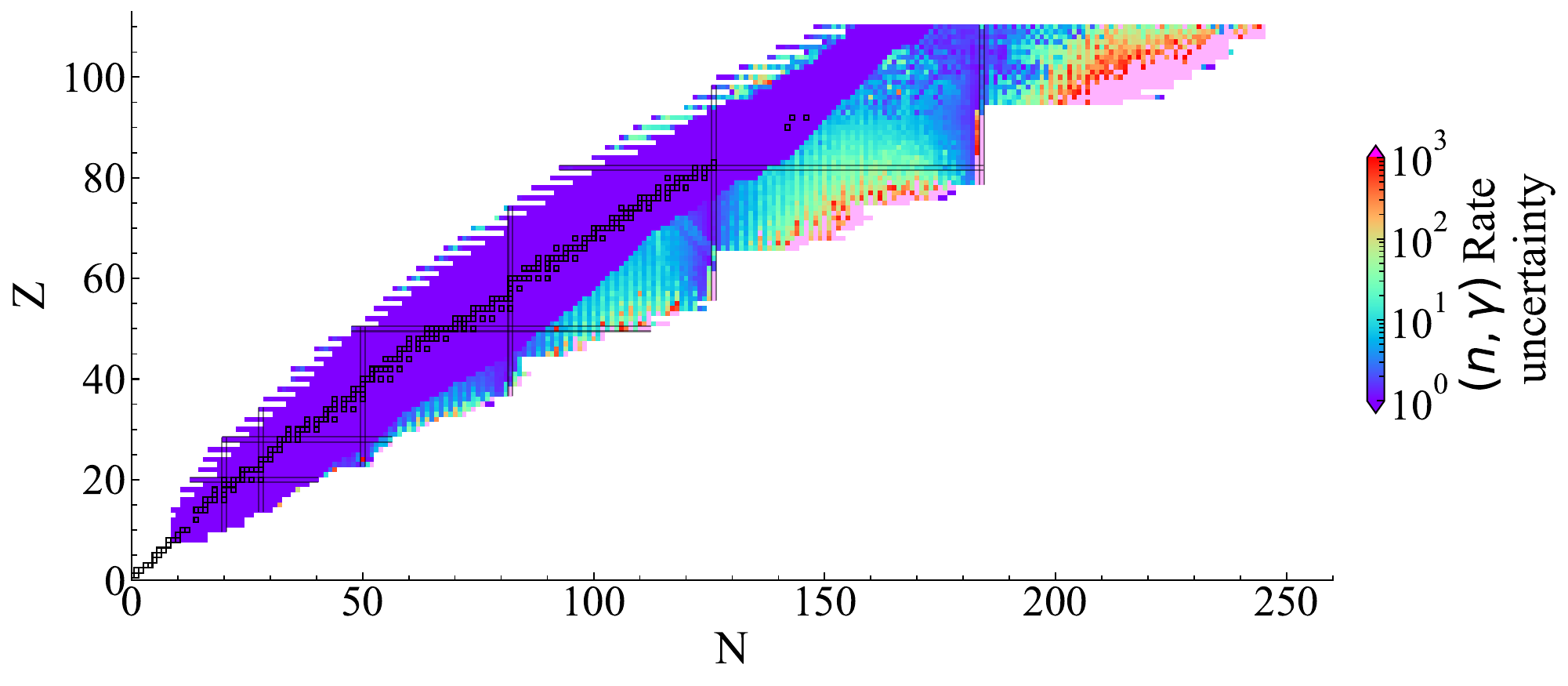}

\caption{{Neutron capture rate uncertainties obtained from TALYS computations using Case 4 mass uncertainties. The reaction are color-coded by their parameter uncertainty (the ratio between maximum and minimum rates). 
}
} 
\label{fig:neutron_capture_rates}
\end{figure*}

\begin{figure*}
    \centering    
\includegraphics[width=0.98\textwidth]{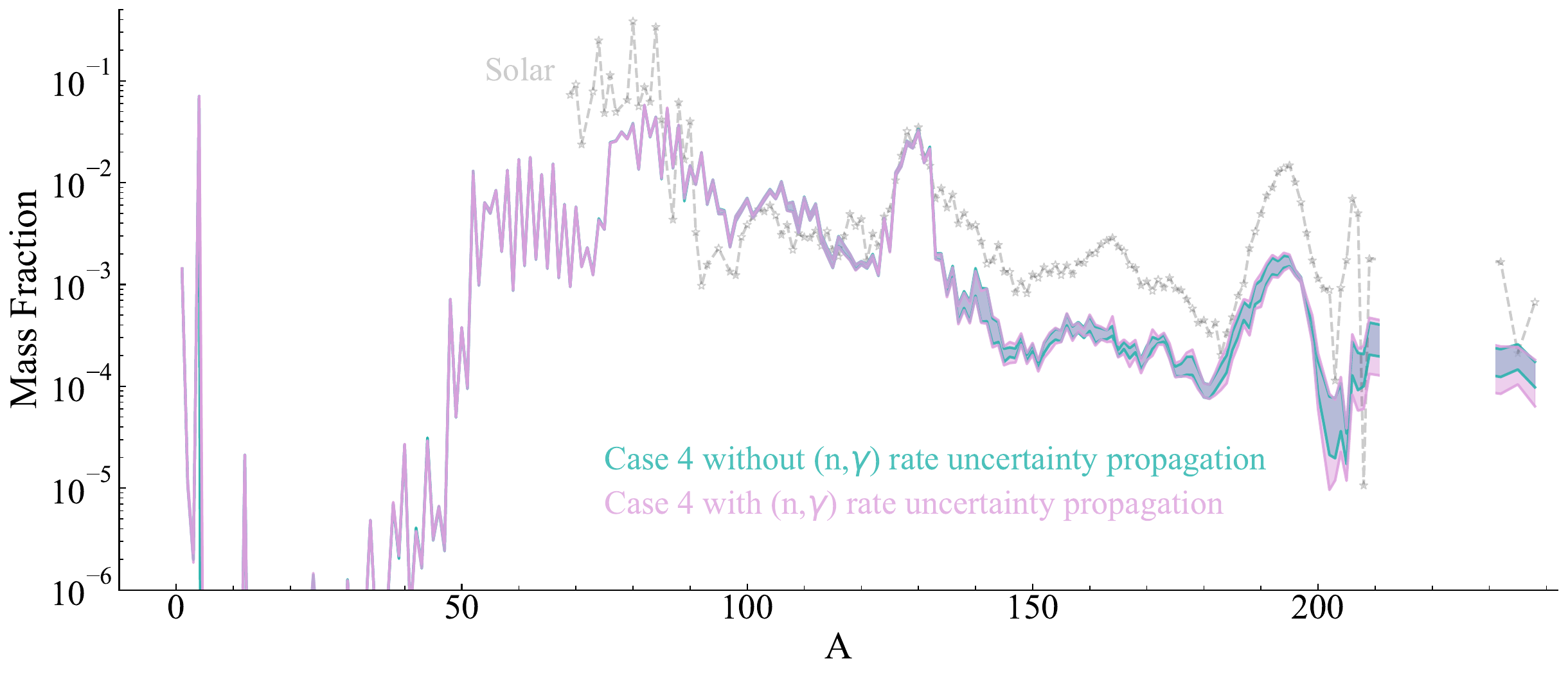}

\caption{{Abundances uncertainties for Case 4 with and without propagation of the mass uncertainties to Hauser-Feshbach computations of neutron capture rates.
}
} 
\label{fig:abundances_with_neutron_capture_rates}
\end{figure*}

So far, we did not propagate the mass uncertainties to the neutron capture rate themselves due to the heavy computation time it entails. However, in the specific Case 4, the number combinations is small and the calculation of reaction rates feasible. The neutron capture rates in Case 4 mass uncertainties were consequently calculated coherently using the TALYS reaction code \citep{Koning23} with the associated masses. These uncertainties are then propagated to r-process nucleosynthesis simulations together with masses for the calculation of photorates. 

Figure \ref{fig:neutron_capture_rates} shows the ratio of the maximum to minimum neutron capture rates resulting from the propagation of nuclear mass uncertainties of Case 4. Experimental masses are used when ever available, so that for those nuclei around the valley of stability no uncertainty associated to theoretical masses exists. Straying further away from the stability results however to large uncertainties on the neutron capture rates, with up to 10$^3$ ratio between the maximum and minimum rates. Small uncertainties are found around magic numbers for the reasons discussed in Fig. \ref{fig:sigma_case2}. 

Fig. \ref{fig:abundances_with_neutron_capture_rates} displays the final abundances uncertainties resulting from the Case 4 nuclear mass uncertainties, taking or not into account the propagation to neutron capture rates. The abundance uncertainties are found to be almost identical with or without a coherent calculation of the neutron capture rates due to the fairly well-established $(n,\gamma)-(\gamma,n)$ equilibrium in the specific NSM model considered here. A slightly larger uncertainty for heavy nuclei production above $A \ga 200$ is obtained, though, when including the propagation to $(n,\gamma)$ rates. 
This test shows that our previous results obtained by propagating the nuclear mass uncertainties without computing coherently the neutron capture rates is a good first-order approximation.

}


\section{Conclusions}
\label{sec:conc}


We investigated both the model (systematic) and parameter (statistical) uncertainties associated with theoretical masses and their correlated neutron separation energies and $Q_\beta$ values relevant for the r-process nucleosynthesis.  We considered 6 mass models to estimate the correlated model uncertainties and for one model, namely HFB-24, we applied the BFMC approach to anchor the parameter uncertainties with available experimental data, {\it i.e.} the known masses. Doing so, we obtain for HFB-24 an estimate of the upper and lower limits to the 5000 unknown masses involved in the r-process network. We explore 4 cases to { propagate the mass uncertainties into the neutron separation energies} taking into account various physical considerations. 
Correlated model uncertainties are overall larger than uncorrelated parameter uncertainties, 
but both can differ significantly for the neutron-rich nuclei involved in the r-process.

We determine the impact of these nuclear uncertainties on the r-process nucleosynthesis in a 1.38-1.38M$_\odot$ symmetrical NSM model \citep[from][]{Just23}, taking into account the production in the dynamical component, the NS-torus transient and the subsequent BH-torus remnant. We considered nucleosynthesis calculation for the 358 trajectories representative of ther overall ejecta. Around 40 models were computed for each case using combinations of maximum and minimum nuclear masses or $S_n$ in order to propagate the parameter uncertainties to the r-process abundances.

We find that using the uncorrelated mass uncertainties leads to a systematic overestimate of the final abundance uncertainties. Using the neutron separation energy, the key quantity defining the r-process path, results in a much better constraint on the uncertainties. However, special attention has to be paid to taking the anti-correlation with the isotopic neighbour into account. Additionally, the anticorrelation existing between $Q_\beta$ and $S_n$ also needs to be taken into account. We finally obtain a coherent estimate of the uncertainties in our so-called Case 4, taking into account these physical requirements. For this chosen case, we obtain r-process abundance uncertainties of the order of 20\% up to $A \simeq 130$, 40\% between $A=150$ and 200, and peaks around $A \simeq 140$ and $A \simeq 203$ giving rise to deviations around 100 to 300\%.
The impact of the parameter uncertainties is smaller than the model uncertainties for most of the r-process production, except for $A \simeq 140$ and $A \simeq 202-203$. We however emphasize  the importance of considering the correlations associated with model uncertainties when propagating them to astrophysics simulations and the potential misleading conclusions that can be drawn when neglecting them, in particular overestimating the impact on abundances. { The parameter uncertainties were also propagated to the radiative neutron capture rates which are found to be potentially affected up to a factor of $10^3$, but, for  the NSM model considered here, such uncertainties on the $(n,\gamma)$ rates have a small impact on abundances since an $(n,\gamma)-(\gamma,n)$ equilibrium is fairly well established.  Similarly, the mass uncertainties were propagated into $Q_\beta$-values but not into $\beta$-decay rates. The latter would require a huge computational effort, in particular within the microscopic QRPA framework used nowadays to estimate $\beta$-decay rates. A simplified approach could consist in considering the approximate relation between $\beta$-decay rates known to be proportional to $Q_\beta^5$, so that the uncertainties propagated into $Q_\beta$ could be further extended to the $\beta$-decay rates. In such an approximation, a 1~MeV uncertainty on a $Q_\beta=15$~MeV decay would lead to an approximate change of  the rate by about 40\%. For this reason, these effects have been considered as smaller than those stemming from changes in $S_n$ and their study postponed to a future work.}

A comprehensive comparison with other sensitivity studies can be found in \citet[][Sect. 4.5]{Kullmann23}, hence will not be repeated here. 
Improvements to nuclear mass models are still crucial in reducing uncertainties in the r-process nucleosynthesis.
Determining coherently parameter uncertainties is also key in such a sensitivity analysis. It remains complex to propagate properly such uncertainties to astrophysical observables.

\
\begin{acknowledgements}
SM and SG has received support from the European Union (ChECTEC-INFRA, project no. 101008324). This work was supported by the F.R.S.-FNRS under Grant No IISN 4.4502.19 and by the F.R.S.-FNRS and the Fonds Wetenschappelijk Onderzoek - Vlaanderen (FWO) under the EOS Project Nr O000422 and O022818F. SG is senior F.R.S.-FNRS research associates. The present research benefited from computational resources made available on Lucia, the Tier-1 supercomputer of the Walloon Region, infrastructure funded by the Walloon Region under the grant agreement n°1910247.\
\end{acknowledgements}

\bibliographystyle{aa} 
\bibliography{astro} 

\begin{thebibliography}{41}
\expandafter\ifx\csname natexlab\endcsname\relax\def\natexlab#1{#1}\fi

\bibitem[{Abbott {et~al.}(2017)Abbott, Abbott, Abbott, {et~al.}}]{Abbott17}
Abbott, B., Abbott, R., Abbott, T., {et~al.} 2017, Phys. Rev. Lett., 119, 161101

\bibitem[{Arnould \& Goriely(2020)}]{Arnould20}
Arnould, M. \& Goriely, S. 2020, Prog. Part. Nucl. Phys., 112, 103766

\bibitem[{Arnould {et~al.}(2007)Arnould, Goriely, \& Takahashi}]{Arnould07}
Arnould, M., Goriely, S., \& Takahashi, K. 2007, Phys. Repts., 450, 97

\bibitem[{Audi {et~al.}(2012)Audi, M., A.H, F.G, M, X., \& B}]{Audi2012}
Audi, G., M., W., A.H, W., {et~al.} 2012, Chinese Physics C, 36, 1287

\bibitem[{Batail {et~al.}(2024)Batail, Goriely, P\'eru, Hialire, Davesne, \& Pastore}]{Batail24}
Batail, L., Goriely, S., P\'eru, S., {et~al.} 2024, Phys. Rev. Lett., submitted

\bibitem[{Bauge \& Dossantos-Uzarralde(2011)}]{Bauge11}
Bauge, E. \& Dossantos-Uzarralde, P. 2011, J. Korean Phys. Soc., 59, 1218

\bibitem[{{Bollig} {et~al.}(2021){Bollig}, {Yadav}, {Kresse}, {Janka}, {M{\"u}ller}, \& {Heger}}]{Bollig21}
{Bollig}, R., {Yadav}, N., {Kresse}, D., {et~al.} 2021, ApJ, 915, 28

\bibitem[{Chadwick {et~al.}(2007)Chadwick, Kawano, Talou, Bauge, Hilaire, Dossantos-Uzarralde, Garett, Becker, \& Nelson}]{Chadwick07}
Chadwick, M.~B., Kawano, T., Talou, P., {et~al.} 2007, Nucl. Data Sheets, 108, 2742

\bibitem[{Cowan {et~al.}(2021)Cowan, Sneden, Lawler, Aprahamian, Wiescher, Langanke, Mart{\'\i}nez-Pinedo, \& Thielemann}]{Cowan21}
Cowan, J., Sneden, C., Lawler, J., {et~al.} 2021, Rev. Mod. Phys., 93, 015002

\bibitem[{Gillanders {et~al.}(2022)Gillanders, Smartt, Sim, Bauswein, \& Goriely}]{Gillanders22}
Gillanders, J.~H., Smartt, S.~J., Sim, S.~A., Bauswein, A., \& Goriely, S. 2022, Mon. Not. Roy. Astron. Soc., 515, 631

\bibitem[{Goriely(1999)}]{Goriely99}
Goriely, S. 1999, Astron. Astrophys., 342, 881

\bibitem[{Goriely(2015)}]{Goriely15c}
Goriely, S. 2015, Eur. Phys. J. A, 51, 22

\bibitem[{Goriely \& Arnould(1992)}]{Goriely92}
Goriely, S. \& Arnould, M. 1992, Astron. Astrophys., 262, 73

\bibitem[{Goriely \& Capote(2014)}]{Goriely14}
Goriely, S. \& Capote, R. 2014, Phys. Rev. C, 89, 054318

\bibitem[{Goriely {et~al.}(2013)Goriely, Chamel, \& Pearson}]{Goriely13a}
Goriely, S., Chamel, N., \& Pearson, J. 2013, Phys. Rev. C, 88, 024308

\bibitem[{Goriely {et~al.}(2009{\natexlab{a}})Goriely, Hilaire, Girod, \& P{\'e}ru}]{Goriely09a}
Goriely, S., Hilaire, S., Girod, M., \& P{\'e}ru, S. 2009{\natexlab{a}}, Phys. Rev. Lett., 102, 242501

\bibitem[{Goriely {et~al.}(2009{\natexlab{b}})Goriely, Hilaire, Koning, Sin, \& Capote}]{Goriely09b}
Goriely, S., Hilaire, S., Koning, A., Sin, M., \& Capote, R. 2009{\natexlab{b}}, Phys. Rev. C, 79, 024612

\bibitem[{Goriely {et~al.}(2007)Goriely, Samyn, \& Pearson}]{Goriely07a}
Goriely, S., Samyn, M., \& Pearson, M.~J. 2007, Phys. Rev. C, 75, 064312

\bibitem[{Grams {et~al.}(2023)Grams, Ryssens, Scamps, Goriely, \& Chamel}]{Grams23}
Grams, G., Ryssens, W., Scamps, G., Goriely, S., \& Chamel, N. 2023, Eur. Phys. J. A, 59, 270

\bibitem[{Janka(2017)}]{Janka17}
Janka, H.-T. 2017, Handbook of Supernovae (Springer International Pub. AG), 1095

\bibitem[{Just {et~al.}(2022)Just, Aloy, Obergaulinger, \& Nagataki}]{Just22b}
Just, O., Aloy, M.~A., Obergaulinger, M., \& Nagataki, S. 2022, ApJ. Lett., 934, L30

\bibitem[{Just {et~al.}(2015)Just, Bauswein, Ardevol~Pulpillo, Goriely, \& Janka}]{Just15}
Just, O., Bauswein, A., Ardevol~Pulpillo, R., Goriely, S., \& Janka, H.-T. 2015, MNRAS, 448, 541

\bibitem[{Just {et~al.}(2023)Just, Vijayan, Xiong, Goriely, Soultanis, Bauswein, Guilet, Janka, \& Martinez-Pinedo}]{Just23}
Just, O., Vijayan, V., Xiong, Z., {et~al.} 2023, ApJ. Lett., 951, L12

\bibitem[{Koning {et~al.}(2023)Koning, Hilaire, \& Goriely}]{Koning23}
Koning, A., Hilaire, S., \& Goriely, S. 2023, Eur. Phys. J. A, 59, 131

\bibitem[{Kullmann {et~al.}(2023)Kullmann, Goriely, Just, Bauswein, \& Janka}]{Kullmann23}
Kullmann, I., Goriely, S., Just, O., Bauswein, A., \& Janka, H.-T. 2023, MNRAS, 523, 2551

\bibitem[{Lema{\^\i}tre {et~al.}(2021)Lema{\^\i}tre, Goriely, Bauswein, \& Janka}]{Lemaitre21}
Lema{\^\i}tre, J.-F., Goriely, S., Bauswein, A., \& Janka, H.-T. 2021, Phys. Rev. C, 103, 025806

\bibitem[{Lunney {et~al.}(2003)Lunney, Pearson, \& Thibault}]{Lunney03}
Lunney, D., Pearson, J., \& Thibault, C. 2003, Rev. Mod. Phys., 75, 1021

\bibitem[{Martinet {et~al.}(2024)Martinet, Choplin, Goriely, \& Siess}]{Martinet24}
Martinet, S., Choplin, A., Goriely, S., \& Siess, L. 2024, A\&A, 684, A8

\bibitem[{{Mendoza-Temis} {et~al.}(2015){Mendoza-Temis}, Wu, Langanke, {Mart{\'\i}nez-Pinedo}, Bauswein, \& Janka}]{Mendoza15}
{Mendoza-Temis}, J. D.~J., Wu, M.~R., Langanke, K., {et~al.} 2015, Physical Review C, 92, 1

\bibitem[{M\"oller {et~al.}(2016)M\"oller, Sierk, Ichikawa, \& Sagawa}]{Moller16}
M\"oller, P., Sierk, A., Ichikawa, T., \& Sagawa, H. 2016, At. Data Nucl. Data Tables, 109-110, 1

\bibitem[{Mumpower {et~al.}(2016)Mumpower, Surman, McLaughlin, \& Aprahamian}]{Mumpower16}
Mumpower, M.~R., Surman, R., McLaughlin, G.~C., \& Aprahamian, A. 2016, Prog. Part. Nucl. Phys., 86, 86

\bibitem[{Nishimura {et~al.}(2015)Nishimura, Takiwaki, \& Thielemann}]{Nishimura15}
Nishimura, N., Takiwaki, T., \& Thielemann, F.-K. 2015, ApJ, 810, 109

\bibitem[{{Pearson}(2001)}]{Pearson00}
{Pearson}, J.~M. 2001, Hyp. Int., 132, 59

\bibitem[{Shen {et~al.}(2015)Shen, Cooke, Ramirez-Ruiz, Madau, Mayer, \& Guedes}]{Shen15}
Shen, S., Cooke, R., Ramirez-Ruiz, E., {et~al.} 2015, ApJ, 807, 115

\bibitem[{Siegel {et~al.}(2019)Siegel, Barnes, \& Metzger}]{Siegel19b}
Siegel, D., Barnes, J., \& Metzger, B. 2019, Nature, 569, 241

\bibitem[{Sprouse {et~al.}(2020)Sprouse, Perez, Surman, Mumpower, McLaughlin, \& Schunck}]{Sprouse20}
Sprouse, T.~M., Perez, R.~N., Surman, R., {et~al.} 2020, Phys. Rev. C, 101, 055803

\bibitem[{Wanajo {et~al.}(2018)Wanajo, M\"uller, Janka, \& Heger}]{Wanajo18a}
Wanajo, S., M\"uller, B., Janka, H.-T., \& Heger, A. 2018, ApJ, 853, 40

\bibitem[{Wang {et~al.}(2021)Wang, Huang, Kondev, Audi, \& Naimi}]{Wang21}
Wang, M., Huang, W., Kondev, F., Audi, G., \& Naimi, S. 2021, Chinese Physics C, 45, 030003

\bibitem[{Wang {et~al.}(2014)Wang, Liu, Wu, \& Meng}]{Wang14}
Wang, N., Liu, M., Wu, X., \& Meng, J. 2014, Phys. Lett. B, 734, 215

\bibitem[{Watson {et~al.}(2019)Watson, Hansen, Selsing, Koch, Malesani, Anja C.~Andersen, Arcones, Bauswein, Covino, Grado, Heintz, Hunt, Kouveliotou, Leloudas, Levan, Mazzali, \& Pian}]{Watson19}
Watson, D., Hansen, C.~J., Selsing, J., {et~al.} 2019, Nature, 574, 497

\bibitem[{Xu {et~al.}(2013)Xu, Goriely, Jorissen, Chen, \& Arnould}]{Xu13}
Xu, Y., Goriely, S., Jorissen, A., Chen, G., \& Arnould, M. 2013, A\&A, 549, 10

\end{thebibliography}

\appendix

\section{Separation energies $S_n$}
Figure \ref{fig:dripline} shows the separation energy $S_n$ as a function of $N$ for 80 isotopic chains with $21 \le Z \le 100$.
\begin{figure*}
    \centering
    \includegraphics[width=0.98\textwidth]{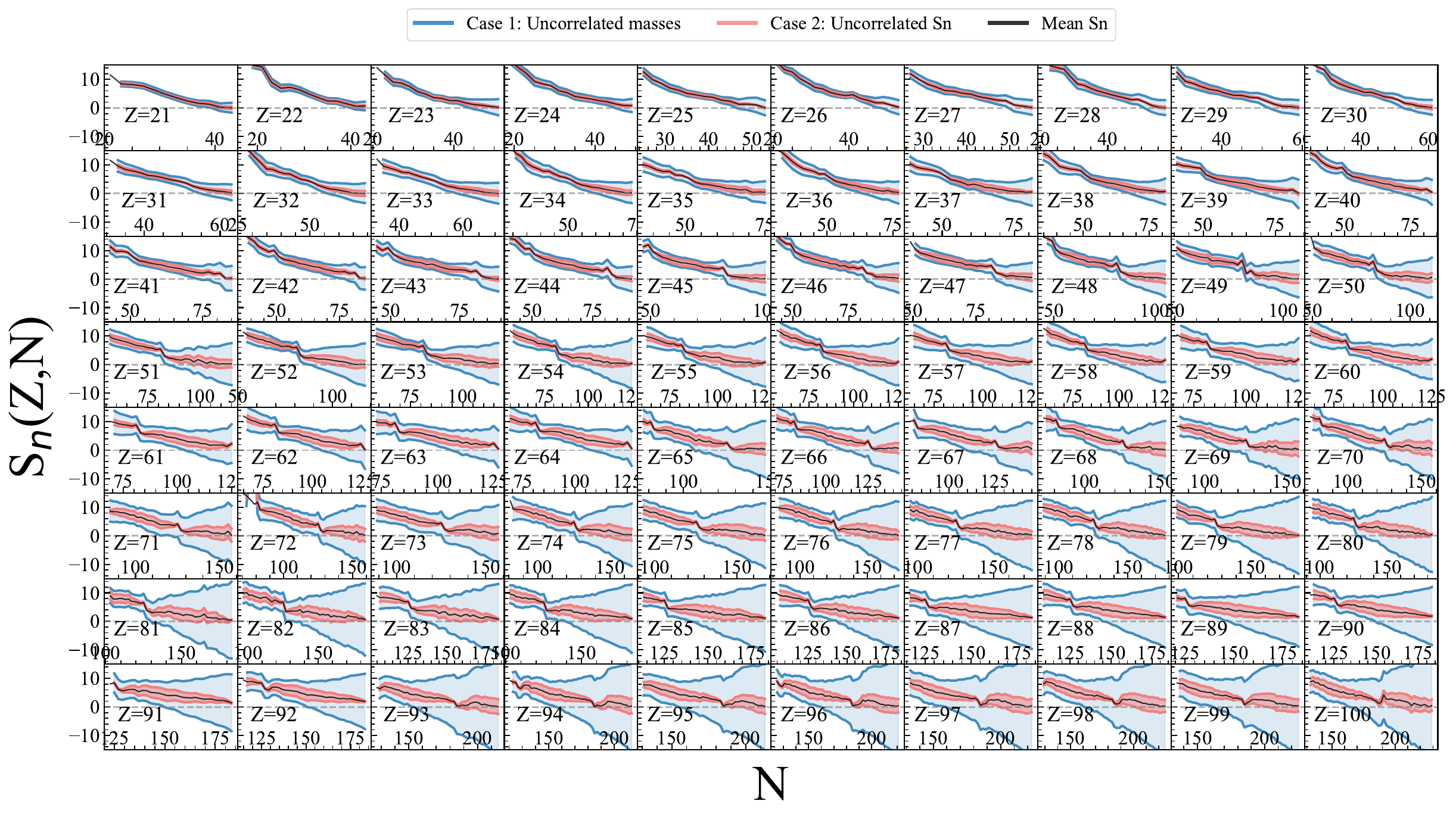}
\caption{{Separation energy $S_n$ as a function of $N$ for 80 isotopic chains with $21 \le Z \le 100$. In blue, $S_n$ maximizing and minimizing the uncorrelated mass uncertainties (Case 1). In red, $S_n$ computed by maximizing and minimizing the uncorrelated $S_n$ uncertainties (Case 2). In black, the mean of the uncorrelated $S_n$ uncertainties.} 
}
\label{fig:dripline}
\end{figure*}

\end{document}